\documentclass[intlimits,twoside,a4paper]{article}

\usepackage{amsmath,amssymb}
\usepackage{graphicx}
\usepackage{siunitx}

\usepackage[T2A]{fontenc}
\usepackage[cp1251]{inputenc}

\usepackage[eqsecnum]{cmpj2}

\issue{2013}{16}{1}{13702}

\doinumber{10.5488/CMP.16.13702}

\title[Calculation of Li--P co-doped diamond]%
{First principles calculation of lithium-phosphorus co-doped diamond%
}
\author[Q.Y. Shao \textsl{et al.}]{Q.Y. Shao\refaddr{label1,label2}\thanks{E-mail: qyshao@163.com, Phone:
+86-20-39310066, Fax: +86-20-39310882}\;,
        G.W. Wang\refaddr{label1}, J. Zhang\refaddr{label1}, K.G. Zhu\refaddr{label3}}
\addresses{
\addr{label1} Laboratory of Quantum Information Technology, School of Physics and Telecommunication Engineering, \\ South China Normal University, Guangzhou 510006, China
\addr{label2} Department of Physics, Zhangzhou Normal University, Zhangzhou 363000, China
\addr{label3} Department of Physics, Beihang University, Beijing 100191, China
}

\date{Received  November 27, 2011, in final form June 30, 2012}
\authorcopyright{Q.Y. Shao, G.W. Wang, J. Zhang, K.G. Zhu, 2013}

\begin{document}

\maketitle

\begin{abstract}
We calculate the density of states (DOS) and the Mulliken
population of the diamond and the co-doped diamonds with different
concentrations of lithium (Li) and phosphorus (P) by the method of the
density functional theory, and analyze the bonding situations of the Li--P
co-doped diamond thin films and the impacts of the Li--P co-doping on the
diamond conductivities. The results show that
the Li--P atoms can promote the split of the diamond energy band near the
Fermi level, and improve the electron conductivities of the Li--P co-doped
diamond thin films, or even make the Li--P co-doped diamond from
semiconductor to conductor. The effect of Li--P co-doping concentration on
the orbital charge distributions, bond lengths and bond populations is
analyzed. The Li atom may promote the split of the energy band
near the Fermi level as well as may favorably regulate the diamond lattice
distortion and expansion caused by the P atom.
\keywords Li--P co-doped diamond, density of states, impurity level, orbital charge
\pacs 71.15.Mb, 71.20.-b, 81.05.Uw, 71.55.Cn
\end{abstract}

\section{Introduction}

Diamond has a great potential for applications. It has a wide band
gap, high breakdown voltage, high carrier mobility, high thermal
conductivity and chemical inertness and so on. By incorporating the
boron atoms into a diamond, the diamond thin films can acquire
shallow acceptor impurity levels near the top of its valence band,
and become the $p$-type semiconductors. Recently, the $p$-type
diamonds have received a lot of developments, and have had a wide range
of applications in the industry, such as the applications of
electrodes~\cite{Lia09}, the detection applications~\cite{Luo09,And10} and the preparations of semiconductor devices~\cite{Yua10}, and so on. In the last two years, the superconducting
characteristics of the boron-doped diamond thin films have attracted
a great deal of attention~\cite{Sop09,Man10,Oki10}. However, the
preparations of the N-type diamond thin films did not have much
progress, which greatly restricted the developments of the diamond
semiconductor devices. N-type impurity atoms are mainly lithium
atoms and the sodium atoms in group I of the Periodic Table, the
nitrogen atoms and the phosphorus atoms in group V, and the
sulfur atoms and the oxygen atoms in group VI. In 1996, by
theoretical calculations, R.~Jones found that the phosphorus vacancy
complexes were deep acceptors, and were capable of compensating  any donor
and making the phosphorus-doped diamond remain an insulator~\cite{Jon96}. In 2005, the thermal ionization energy and the capture
cross-section of the phosphorus donor were estimated to 0.54$\pm
$0.02~eV and (4.5$\pm $2.0)$\times $10$^{-17}$~cm$^{2}$ in the
experiment which analyzed the conductivity of a phosphorus-doped
diamond~\cite{Koi05}. The electrical properties and the shapes of
phosphorus-doped diamonds, which are prepared by an organic
phosphorus gas [P(C$_{4}$H$_{9}$)H$_{2}$ and P(CH$_{3}$)$_{3}$] or
by an inorganic phosphorus gas PH$_{3}$, are the same in the
experiment of the preparation of a phosphorus-doped diamond by the
plasma enhanced chemical vapor deposition method~\cite{Kat05}. At
the same time, in the analysis of the relationship between the
mobility of the phosphorus-doped diamond's (001) surface and
temperature, the largest mobility (approximately 450~cm$^{2}$/Vs)
lies in 260~K, and the mobility at room temperature is 350~cm$^{2}$/Vs~\cite{Kat07}. In 2006, when Takatoshi Yamada and his
collaborators studied the field emission properties of heavy
phosphorus-doped diamond thin films, they found that the
reconstruction film surface which was annealed in vacuum, had a minimum
threshold field value 16~V/\si{\micro\meter}, while the threshold field values
of the thin film surface which was terminated by oxygen or
hydrogen were 28~V/\si{\micro\meter} or 44~V/\si{\micro\meter}~\cite{Yam06}. In 2008, J.~Pernot and his collaborators found that at the phosphorus atom
concentrations being  less than 10$^{17}$~cm$^{-3}$, the electron
mobility was determined  by the lattice scattering; when the phosphorus
atom concentrations were between 10$^{17}$~cm$^{-3 }$ and 10$^{18}$~cm$^{-3}$, the electron mobility was determined by both the lattice
scattering and by the scattering of ionized impurity atoms; when the
phosphorus atom concentrations were higher than 10$^{18}$~cm$^{-3}$,
the electron mobility was determined by the scattering of neutral impurity atoms \cite{Per08}. Although both the experiments and theories
of only single phosphorus atoms doped diamond have greatly
increased and improved, the electron conductivities of the
phosphorus-doped diamond thin films remain low in experiments and
cannot meet the requirements for the preparations of
semiconductor devices. Thus, some researchers referred to the doping
experiments of the GaAs and advised that the co-doping method might
be a good way. In 1995, the experiment of the preparation of the
nitrogen-phosphorus co-doped diamond thin films by hot-filament
chemical vapor deposition showed that adding some nitrogen atoms was
advantageous to the phosphorus-doping and to the increase of the growth
rate of the films, as well as can get a higher doping concentration: the maximum
concentration of the phosphorus atoms and the nitrogen atoms were
 $3 \times 10^{19}$ and $6 \times 10^{19}$~atoms/cm$^{3}$, respectively \cite{Cao95}. In 2004, when Li Rong-Bin and his collaborators
prepared the boron-sulfur co-doped diamond thin films by traditional
microwave plasma chemical vapor deposition, they found that the addition of
 some boron atoms could be advantageous to sulfur doping, and
made the amount of the sulfur atoms increase 1.5 times while the
activation energy of the electron conductivity was reduced from 0.52~eV to 0.39~eV~\cite{LiR04}. In 2005, W.S.~Lee and his collaborators found
that the Hall coefficient of the boron-lithium co-doped diamond thin
films was $-2.974\times 10^{-2}$~cm$^{3}$/C, and its resistivity
was $0.01 \div  0.02~\Omega$m. They also confirmed that the
co-doping method could improve the stability of the lithium in the
diamond thin films~\cite{Lee05}. In 2007, E.B.~Lombardi and his
collaborators also studied the interstitial doping and the substitution
doping of the lithium and sodium atoms. By the experiment they
confirmed that the lithium atom was an interstitial atom and the sodium
atom was a substitution atom~\cite{Lom07}. In 2009, according to
the calculation of the first principle, F.~Iori found that the
co-doped diamond, where two kinds of impurity atoms lied in the
nearest neighbor, has the lowest impurity formation energy~\cite{Ior09}.

In recent years, although the co-doping method had been tried a lot in
experiments and theories, it did not get much progress. The studies
of the Li--P co-doped diamonds are very few. Therefore, in this paper, based
on the first principle of the density functional theory (DFT), we calculate
the Mulliken population and the DOS of the co-doped diamonds with different concentrations of Li and P, analyze their electronic structures,
and determine the bonding properties and the charge distributions among
lithium atoms, phosphorus atoms and carbon atoms and the impacts on the
electrical properties after doping.

\begin{figure}[tbp]

\begin{minipage}[htbp]{0.48\textwidth}
\centerline{\includegraphics[height=4cm]{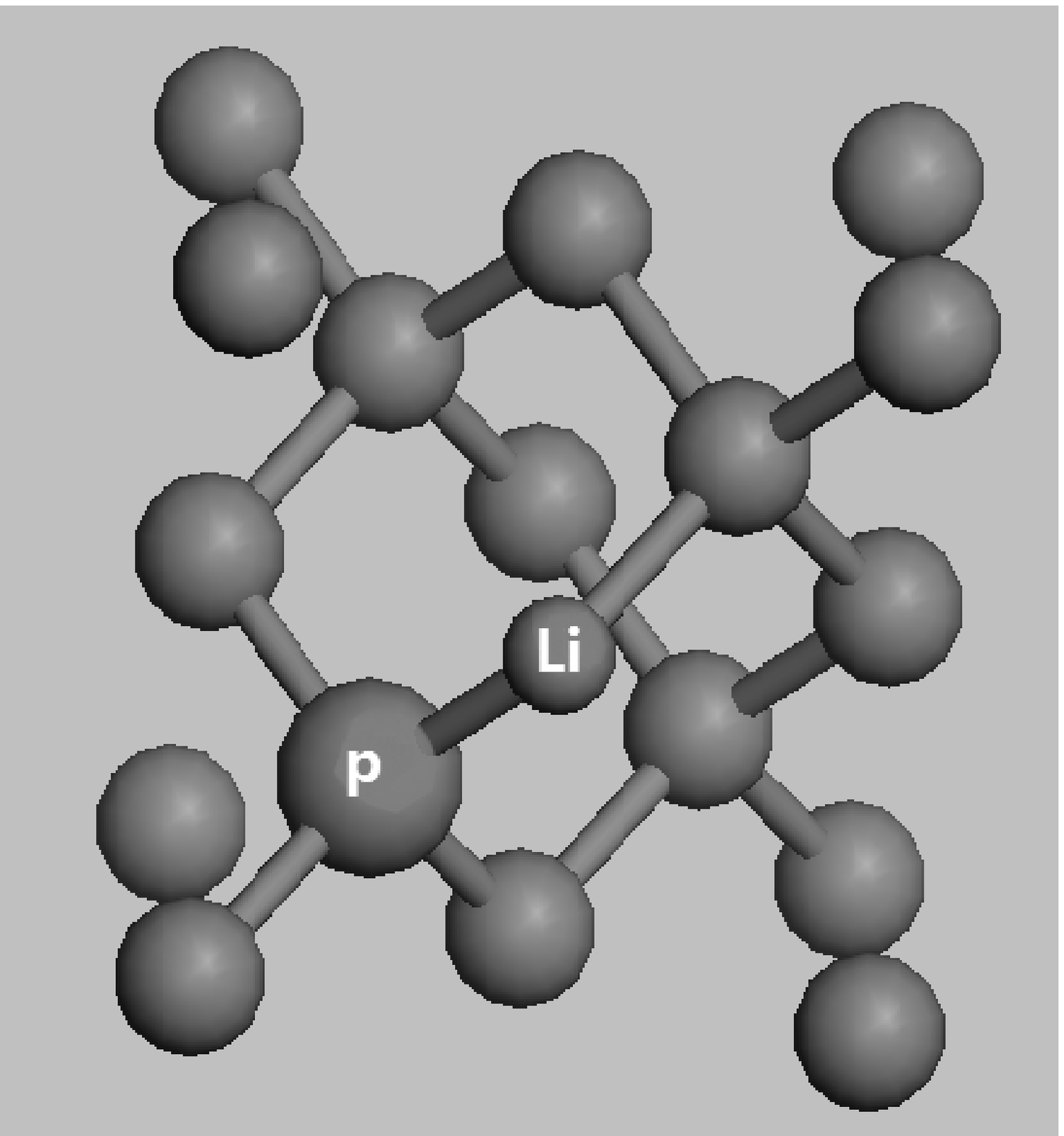}}
\centerline{(a)}
\end{minipage}
\begin{minipage}[htbp]{0.48\textwidth}
\centerline{\includegraphics[height=4cm]{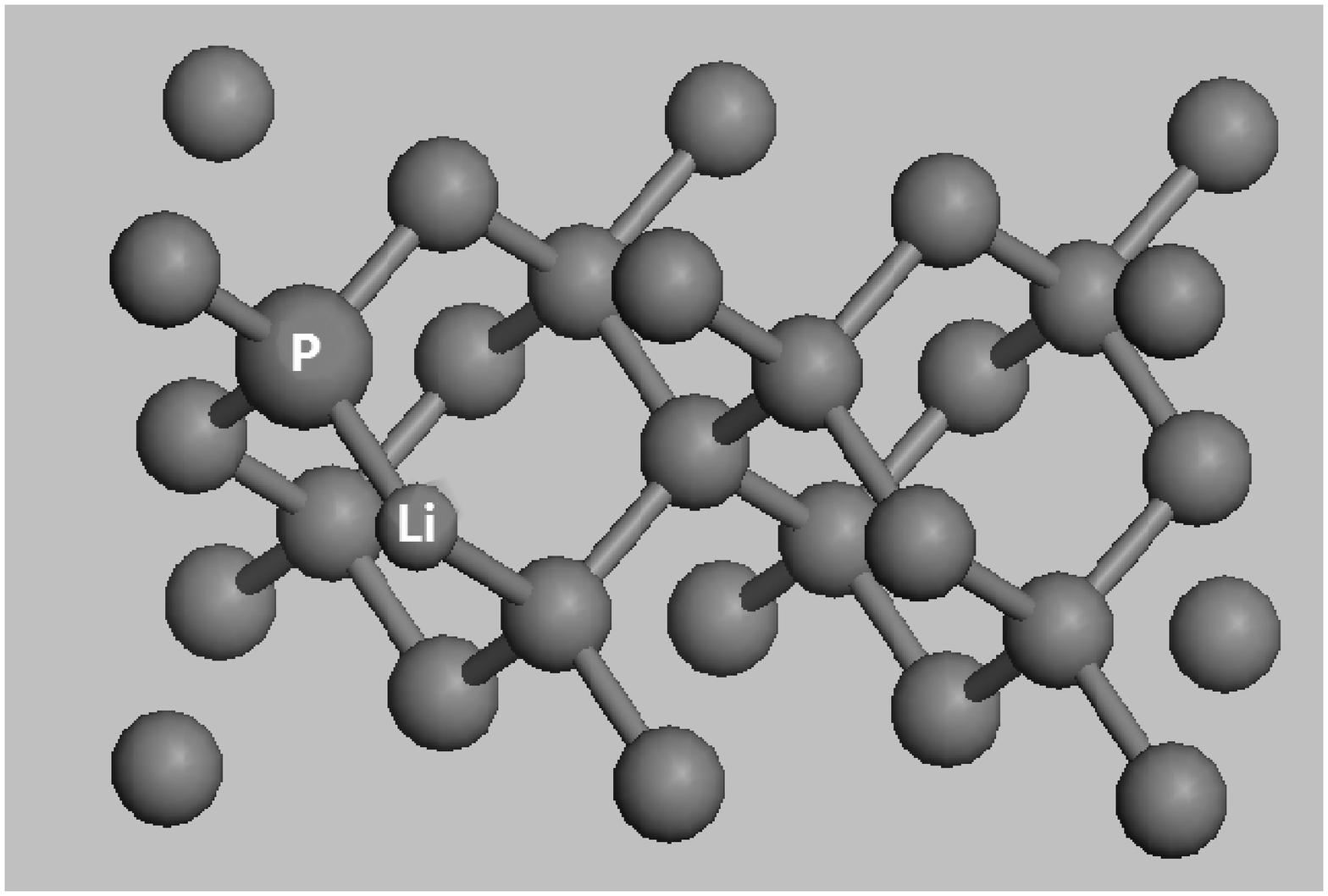}}
\centerline{(b)}
\end{minipage}

\begin{minipage}[htbp]{0.48\textwidth}
\centerline{\includegraphics[height=5cm]{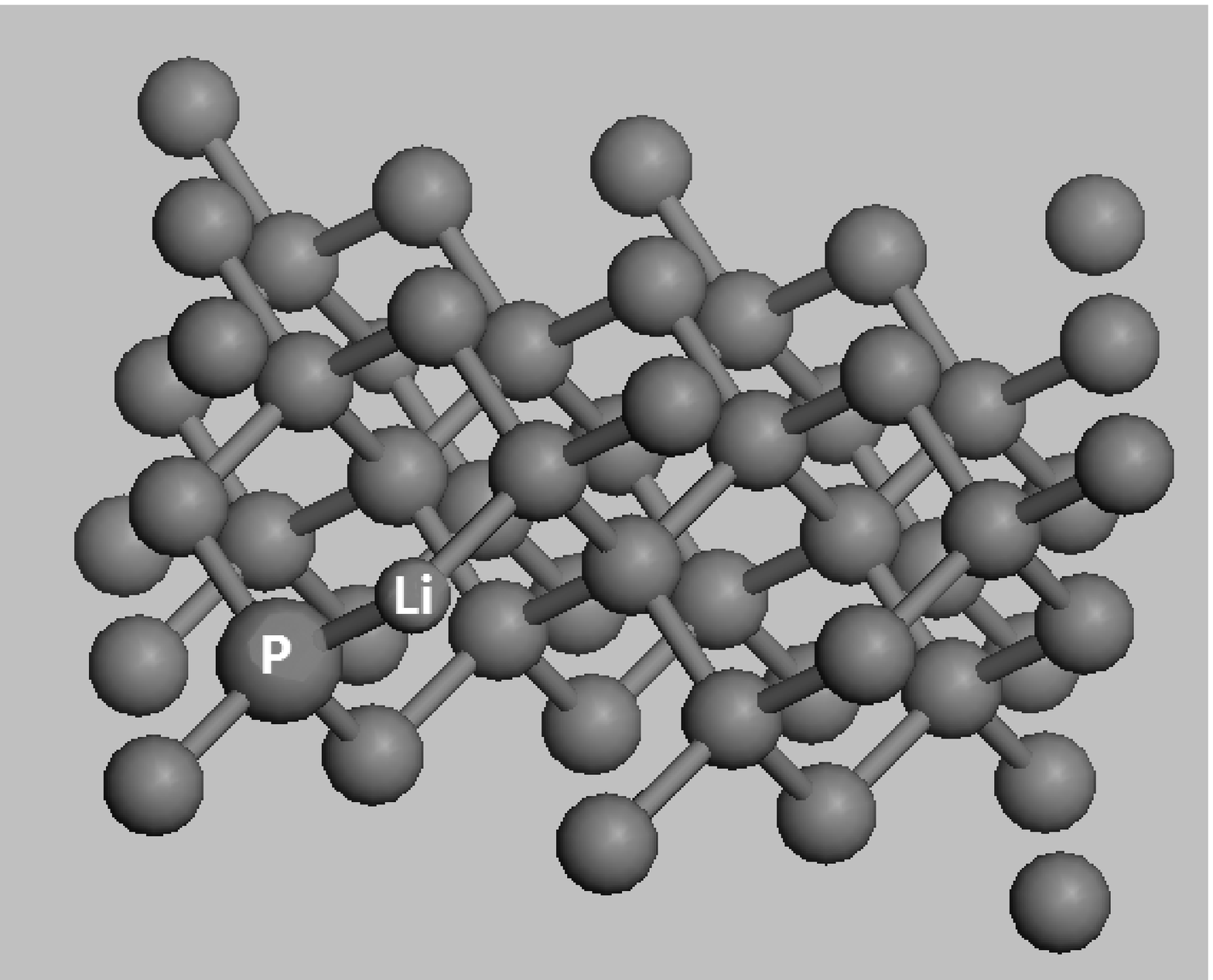}}
\centerline{(c)}
\end{minipage}
\begin{minipage}[htbp]{0.48\textwidth}
\centerline{\includegraphics[height=5cm]{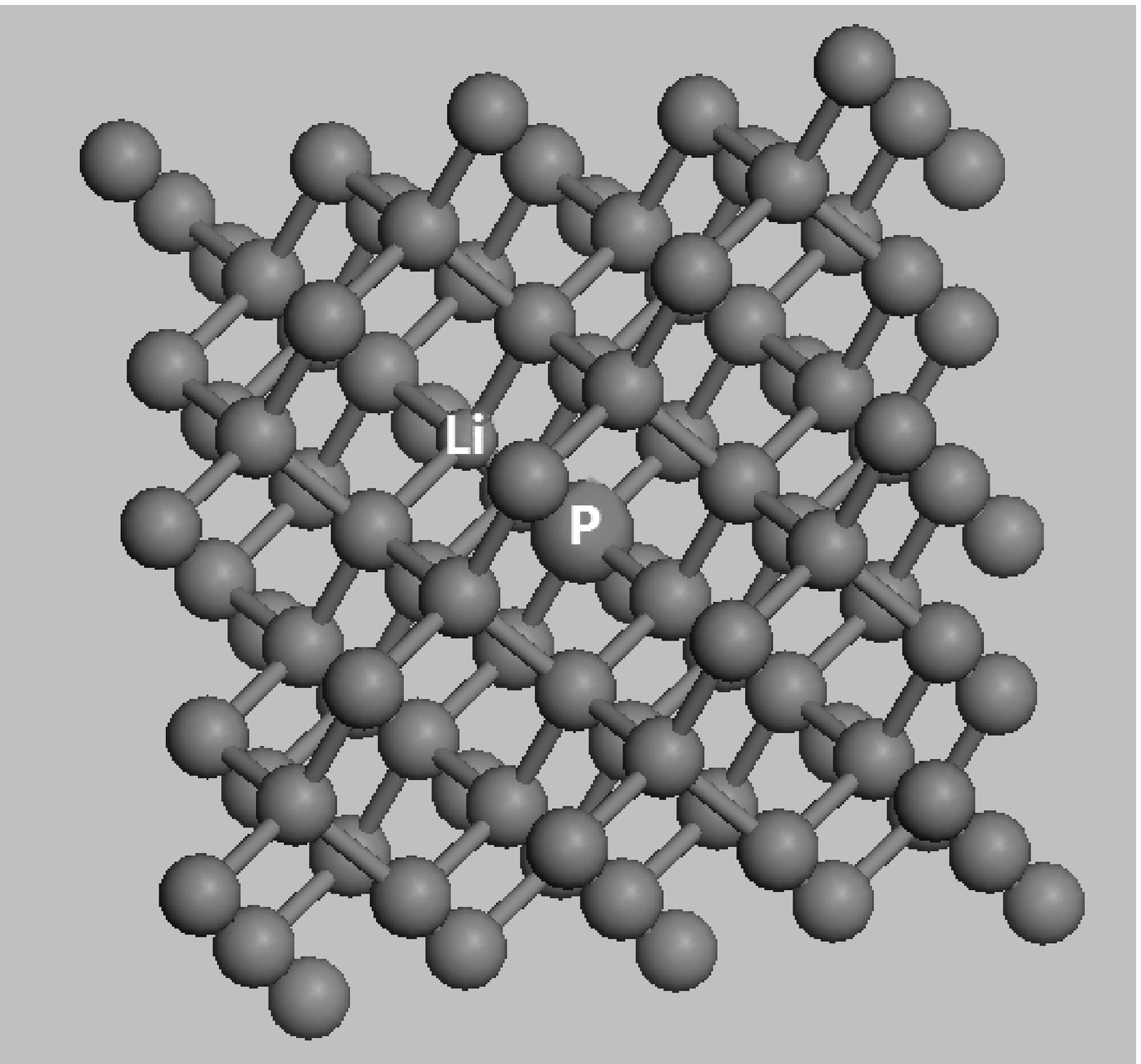}}
\centerline{(d)}
\end{minipage}

\begin{minipage}[htbp]{0.48\textwidth}
\centerline{\includegraphics[height=5.1cm]{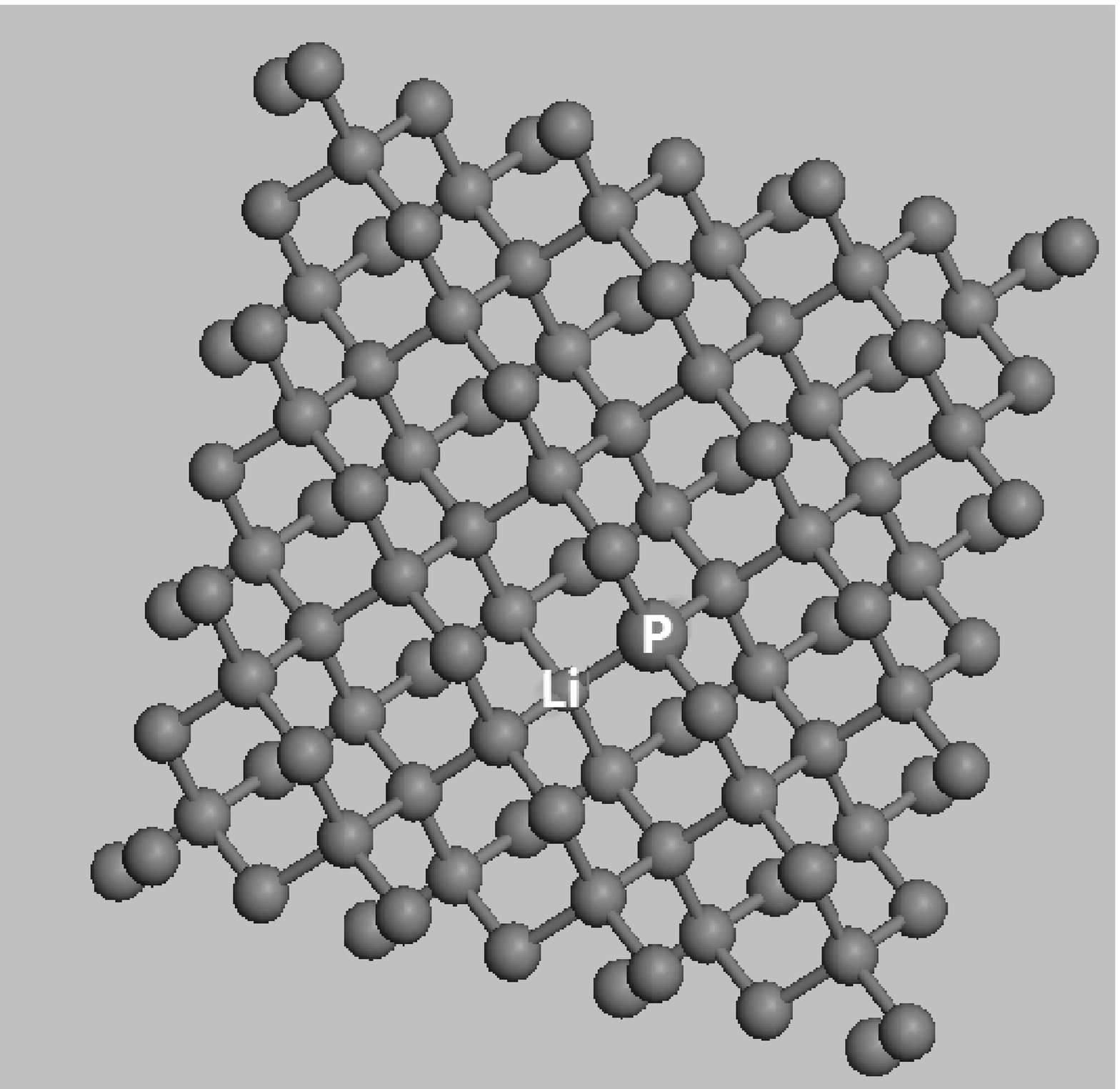}}
\centerline{(e)}
\end{minipage}
\begin{minipage}[htbp]{0.48\textwidth}
\centerline{\includegraphics[height=5.1cm]{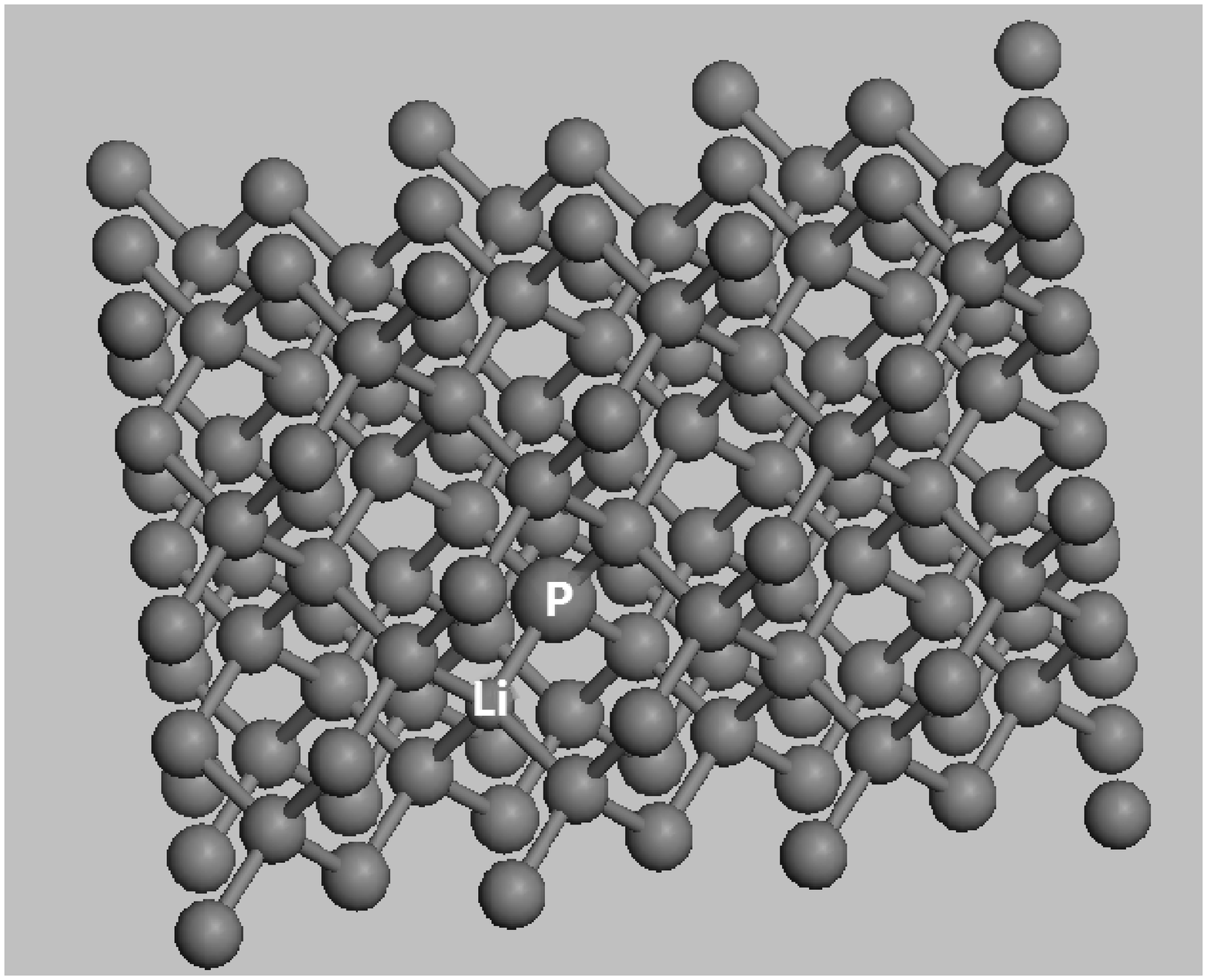}}
\centerline{(f)}
\end{minipage}

\begin{minipage}[htbp]{0.48\textwidth}
\centerline{\includegraphics[height=5.3cm]{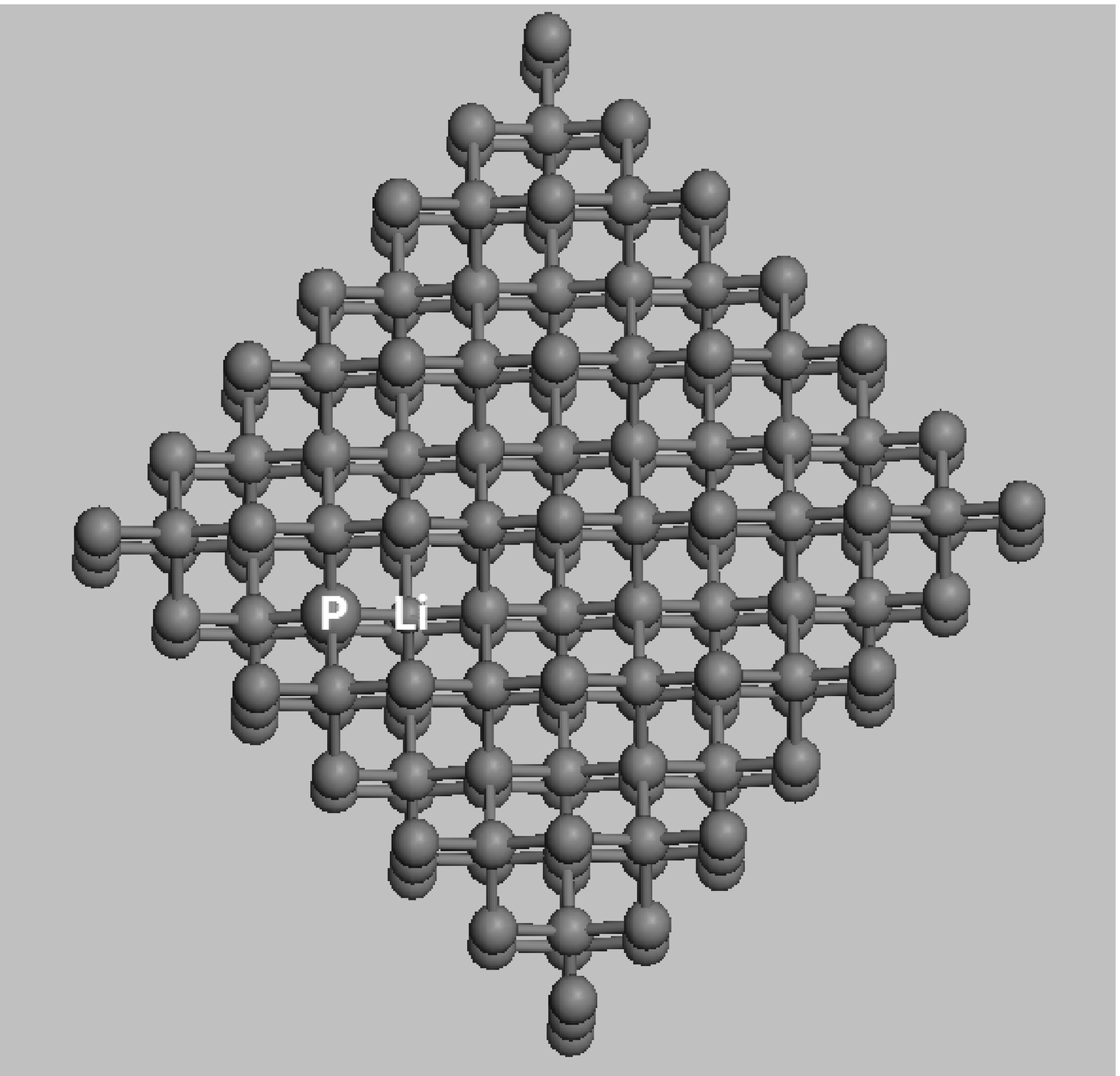}}
\centerline{(g)}
\end{minipage}
\begin{minipage}[htbp]{0.48\textwidth}
\centerline{\includegraphics[height=5.3cm]{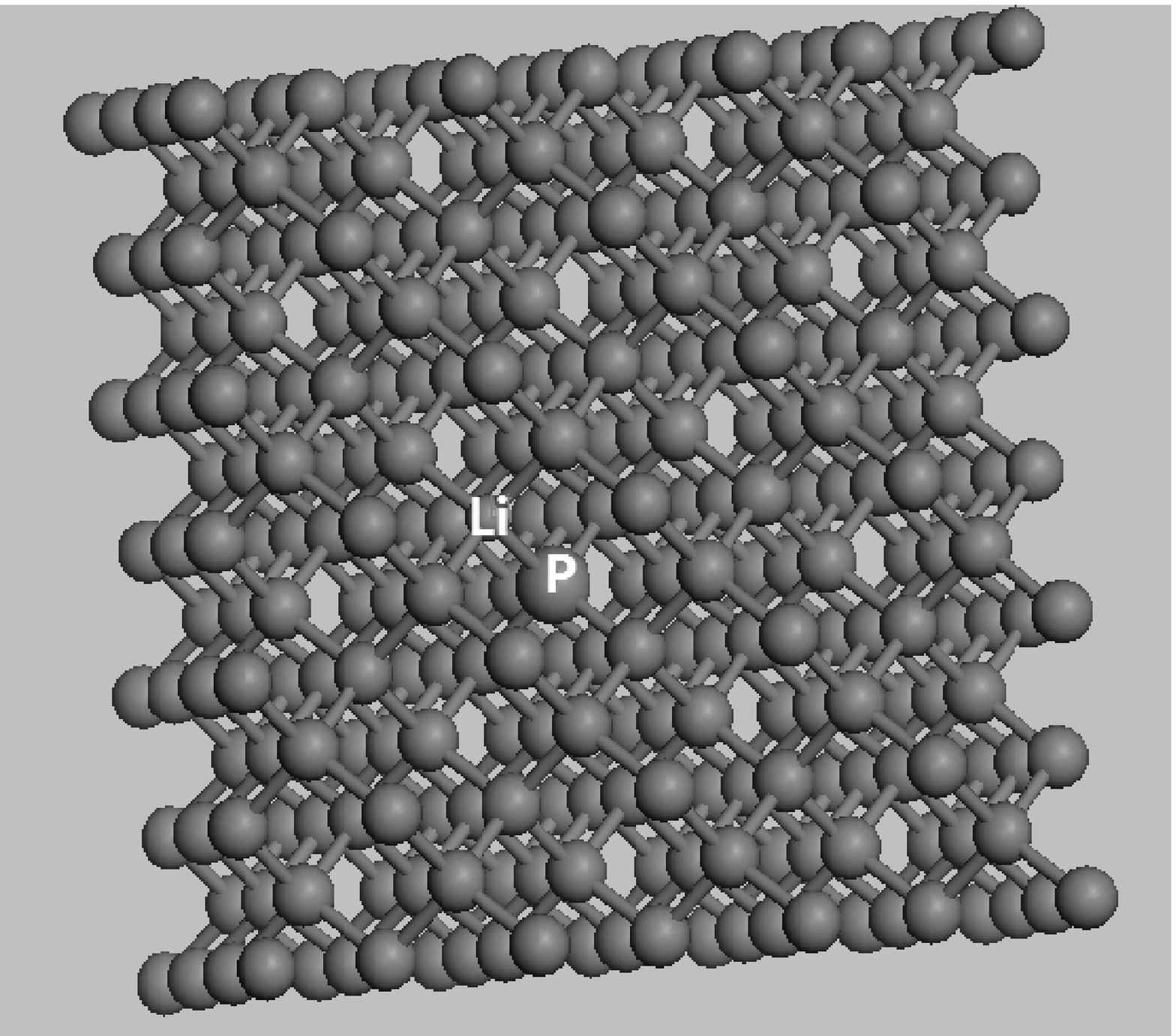}}
\centerline{(h)}
\end{minipage}
\caption{The structure of the doped diamond where the number
of Li atom is one, P atom's number is one and the C atom's number is (a) six, (b) fourteen, (c) thirty, (d) sixty-two, (e) seventy, (f) ninety-four, (g) one-hundred-forty-two, (h) two-hundred-fourteen.}
\label{fig1}
\end{figure}

\section{Calculation method}

In this paper, we do our calculation work with \textit{ab initio} calculation
quantum mechanics module Cambridge Serial Total Energy Package
(CASTEP) which is based on the density functional theory in Accelrys
Materials Studio software~\cite{Seg02}. The module uses  local
atom-based groups and the numerical periodic boundary condition to
describe the valence electrons while the interactions between
electrons and ions is mainly described by the Norm-Conserving
Pseudopotential and the Ultrasoft Pseudopotential~\cite{Van90}. In
order to minimize the number of plane wave basis sets, we chose the
Ultrasoft Pseudopotential to describe the interactions between
electrons and ions, while the valence electron configurations of
the carbon atom, the phosphorus atom, the lithium atom are
respectively selected to be C: $2s^{2}2p^{2}$, P: $3s^{2}3p^{3}$
and Li: $1s^{2}2s^{1}$. The cut off energy of the plane waves is
300.0~eV, and the exchange-correlation energy is described by the
PBE parameterized form of the generalized gradient approximation
(GGA)~\cite{Per96}. In this paper, we only calculate the situation
of two kinds of impurity atoms lying in the nearest neighbor. In
this way, a Li atom and a phosphorus atom are incorporated into the
diamond lattice with an atom pair at a random site. In order to
study the Li--P co-doped diamond lattices of different
concentrations, we separately calculate the diamond lattices where
the ratio of the lithium atom, phosphorus atom, and the carbon atoms
are $1: 1: 6$, $1: 1: 14$, $1: 1: 30$, $1: 1: 62$, $1: 1: 70$, $1: 1: 94$, $1: 1:
142$, $1: 1: 214$, and the structures are shown in figure~\ref{fig1}.
Their $k$ values are set for $7 \times  7 \times  7$, $4 \times 7
\times  7$, $4 \times  4 \times  7$, $4 \times  4 \times 4$, $2
\times 2 \times  7$, $2 \times  4 \times  4$, $2 \times 2
\times 4$, $2 \times  2 \times 2$, in order to ensure the
convergences of the system's energy and the structures in the
plane-wave basis set.

In the SCF calculation, we chose the Pulay density hybrid approach, and set
the SCF to $1.0\times 10^{-6}$~eV$\cdot$atom$^{-1}$. In the geometric
optimization, we chose the BFGS algorithm. The accuracy of total energy was
$1.0\times 10^{-5}$~eV$\cdot$atom$^{-1}$, the crystal force of each atom was
less than 0.3~eV$\cdot$nm$^{-1}$, the stress of each structural unit was less than
0.05~GPa, and the atomic displacement caused by the changes of the
structural parameters was less than $1.0\times 10^{-4}$~nm.

\section{Calculation results and discussions}

\subsection{Analysis of electron density of states of the
lithium-phosphorus co-doped \\ diamond}

\begin{figure}[!h]
\centerline{\includegraphics[width=0.65\textwidth]{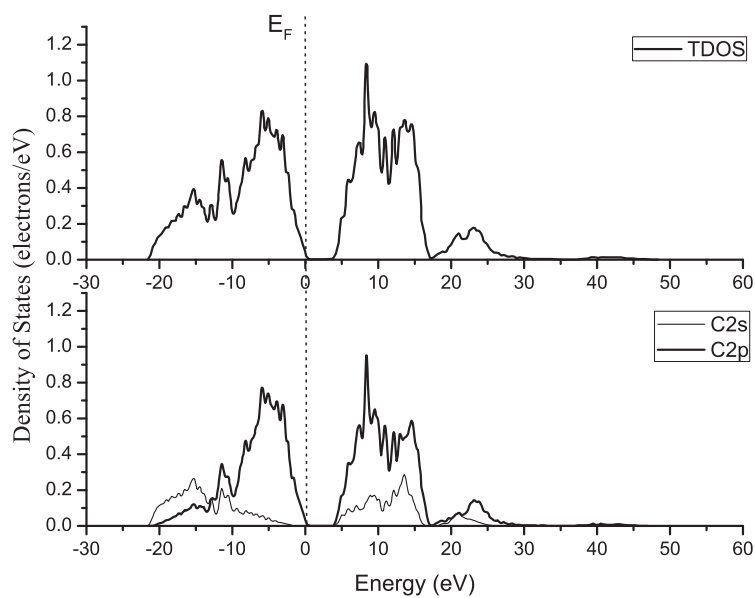}}
\caption{The total density of states (TDOS) and the partial density
of states of the diamond. The $E_\mathrm{F}$ stands for the Fermi level.
The C$2s$ and C$2p$ represent the electric charge distributions of $2s$
and $2p$ orbits.} \label{fig2}
\end{figure}

In this paper, we calculated the DOS of the diamond and the co-doped
diamonds with different concentrations of Li and P. The total density of
states (TDOS) of the diamond and the partial density of states (PDOS) of the
diamond are showed in figure~\ref{fig2}. The middle point line is the Fermi
level. Its left is the valence band and its right is the energy gap and the
conduction band. The energy gap is 4.138~eV and it was a little different from
the value of the experimental measurement 5.4~eV. The phenomenon of the
energy gap being underestimated lies prevalently in the density functional
calculations~\cite{Chi89}, but it does not affect our following qualitative analysis
on the doped diamonds. The valence band of a diamond consists of two
areas: in the high energy region (approximately $-13\div$ 0~eV), it is
mainly occupied by the C$2p$ states, while in the low energy region
(approximately $-21.5\div -13$~eV), it is mainly occupied by the C$2s$ states,
and the conduction band of the diamond is mainly occupied by the C$2p$ states.
From the PDOS in the conduction band, we can see that the area ratio of the
C$2s$ states and C$2p$ states is approximately $1:3$. Here, although the above
result has been known for a long time, we illustrate it again in order to be compared
with the situations of a doped diamond.

\begin{figure}[!t]

\begin{minipage}[htbp]{1.0\linewidth}
\centerline{\includegraphics[width=0.58\textwidth]{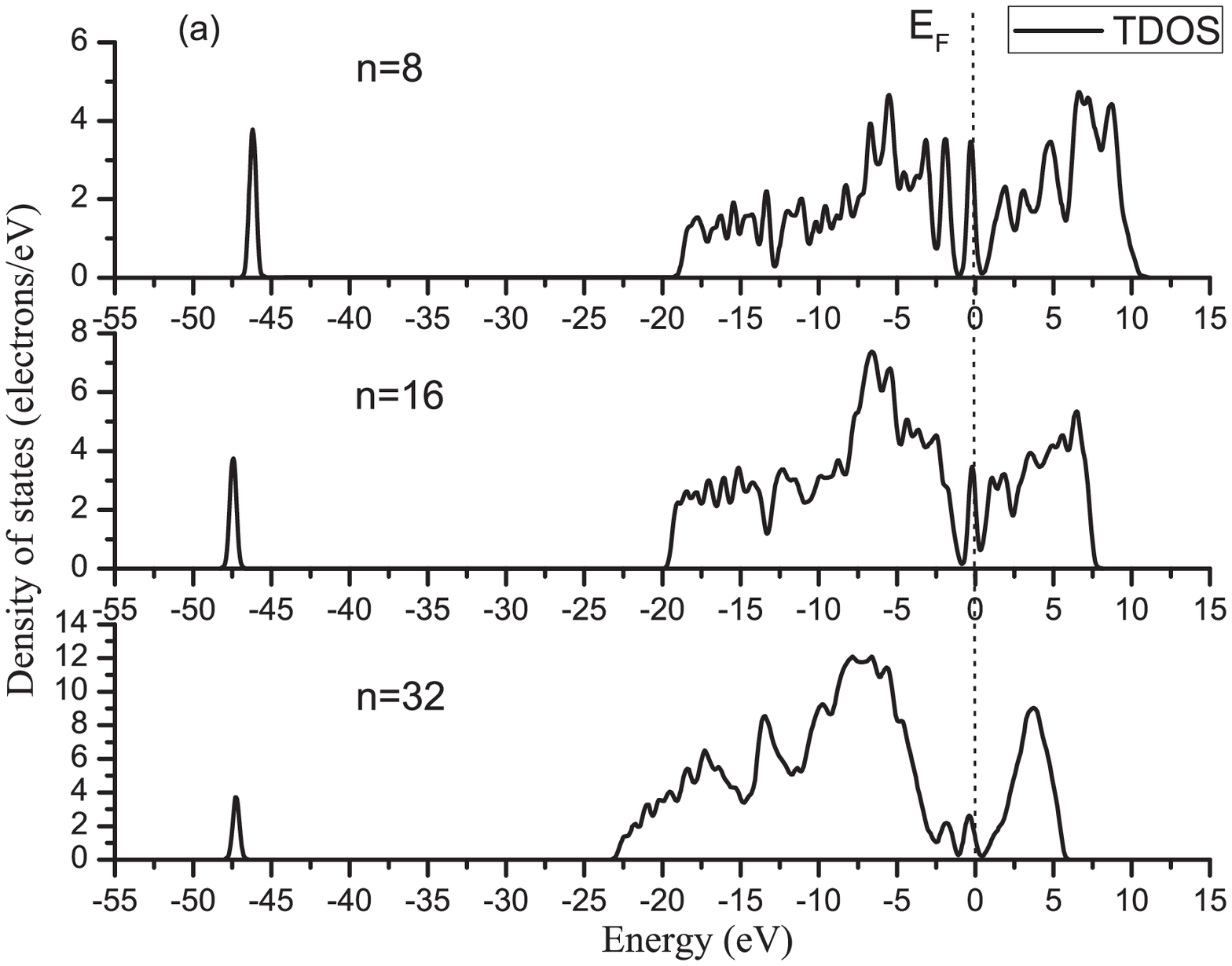}}   
\end{minipage}
\\[5mm]
\begin{minipage}[htbp]{1.0\linewidth}
\centerline{\includegraphics[width=0.62\textwidth]{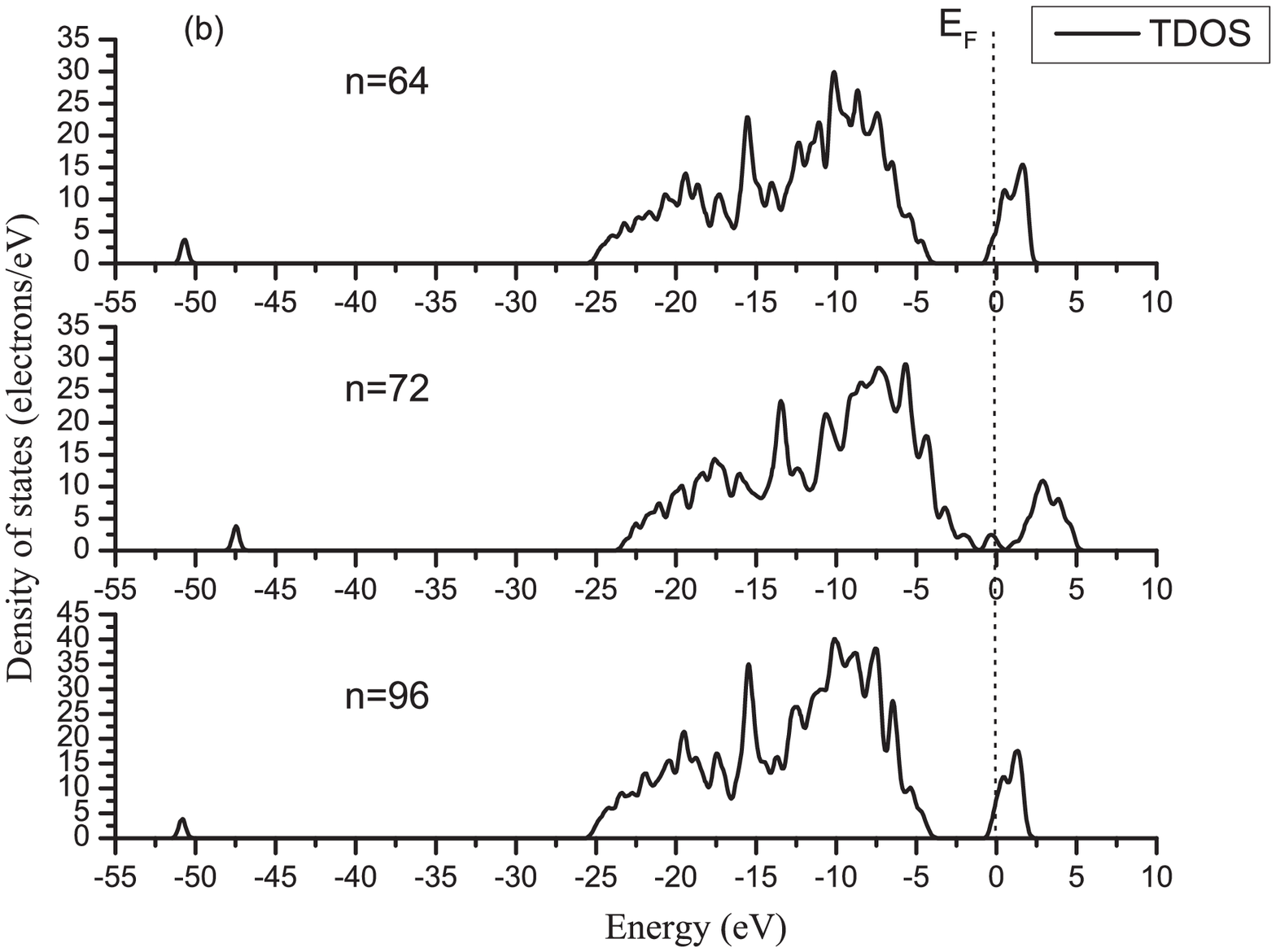}}
\end{minipage}
\\[3mm]
\begin{minipage}[htbp]{1.0\linewidth}
\centerline{\includegraphics[width=0.61\textwidth]{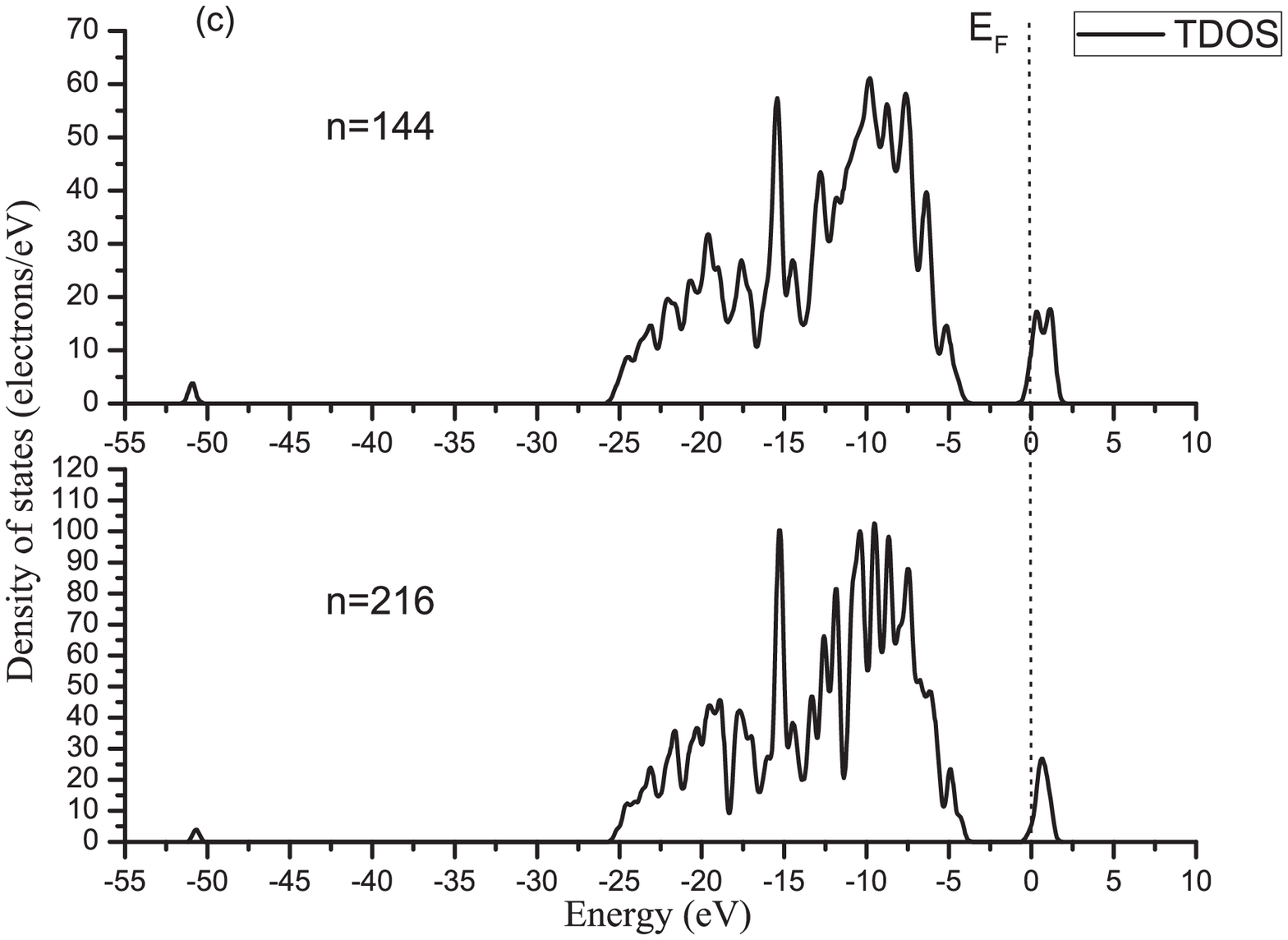}}
\end{minipage}
\caption{The total density of states (TDOS) of the Li--P co-doped
diamond, $n$ is the total atomic number, and each cell has a
lithium atom and a phosphorus atom (the Li--P atom-pair). The $E_\mathrm{F}$
stands for the Fermi level.} \label{fig3}
\end{figure}

\clearpage

Comparing the TDOS of the Li--P co-doped diamond in figure~\ref{fig3} with the TDOS of
the diamond in figure~\ref{fig2}, we can see that the Fermi level of a diamond after
doping obviously moves near the bottom of conduction band. When the doping
concentration is relatively low, such as the TDOS shown in figure~\ref{fig3}~(b)
and~(c), the Fermi level of the Li--P co-doped diamond moving near the bottom
of the conduction band, their energy gap width is almost the same as the
diamond's (as shown in figure~\ref{fig2}), and the donor levels are formed in the
band gap and near the bottom of the conduction band. However, when the doping
concentration is higher, such as the TDOS shown in figure~\ref{fig3}~(a), the
energy gap of the Li--P co-doped diamond disappears. This is most
probably caused by the impurity level broadening and an impurity band
formation, which completely fills the energy gap region. Thus, the Li--P
co-doped diamond becomes a conductor. All this illustrates that when the
donor impurity atoms are incorporated into a diamond, the impurity levels
will appear near the bottom of the conduction band, and then the insulated
diamond will become a semiconductor which has some conductivity. When
the concentrations of the lithium atom and phosphorus atom are not too high,
only some impurity levels are formed near the bottom of the conduction band. But
when the concentrations of the lithium atom and phosphorus atom are high
enough, the impurity levels will increase, broaden and become an impurity
band, or even extend to the entire energy gap. Then, they will make the
energy gap disappear, and the semiconductor will become a conductor. Here,
the magnitude of concentration of the lithium atom or the phosphorus
atom is about $10^{21}$, three orders of magnitude higher than the
concentration of the phosphorus atoms in the reference~\cite{Kat05} where its
concentration is about $(2 \div 3)\times 10^{18}$~cm$^{-3}$, and two
orders of magnitude higher than the concentration of the phosphorus atoms in
the reference~\cite{Yam06} where its concentration is determined to be $7\times 10^{19}$~cm$^{-3}$. Although some impurity atom concentrations of the
calculation models in this paper are higher than in the experiment, the
change or tendency of the doped diamond for different doping concentrations
in this paper is clear, so it does not affect our analysis.
\begin{figure}[!b]
\centerline{\includegraphics[width=0.65\textwidth]{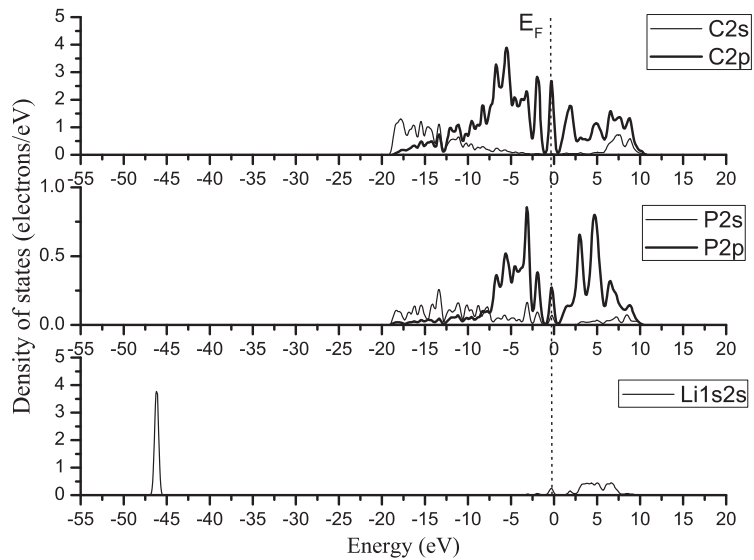}}
\caption{The partial density of states of the doped diamond where
the number of Li atom is one, P atom's number is one and the C
atom's number is six. The $E_\mathrm{F}$ stands for the Fermi level. The
C$2s$ and C$2p$ represent the electric charge distributions of $2s$ and $2p$
orbits of the carbon atoms. The P$2s$ and P$2p$ represent the electric
charge distributions of $2s$ and $2p$ orbits of the phosphorus atoms.
The Li$1s2s$ represent the electric charge distributions of $1s$ and $2s$
orbits of the lithium atoms.} \label{fig4}
\end{figure}
In addition, we
also found that in the TDOS of the doped diamond whose energy gap has
disappeared (as shown in figure~\ref{fig3}), the peak value of the DOS of the valence band of a lithium
atom lies between about $-48.5\div  -45$~eV, while in the
TDOS of the doped diamond whose energy gap has not disappeared, the peak
value of the DOS of the valence band of a lithium atom lies between about
$-51.5\div  -50$~eV. This indicates that when the doping concentration of the
Li--P co-doped diamond is high, the DOS of the valence band of a lithium atom
will move toward the high energy direction. In short, when the atomic
ratio of the lithium atoms or the phosphorus atoms to the carbon atoms is
less than $1:70$ (or the concentration of the lithium atom or the phosphorus
atom is less than $2.35 \times 10^{21}$~cm$^{-3})$, the Li--P co-doped
diamonds are always semiconductors, otherwise they are
conductors.

\begin{figure}[!t]
\centerline{\includegraphics[width=0.65\textwidth]{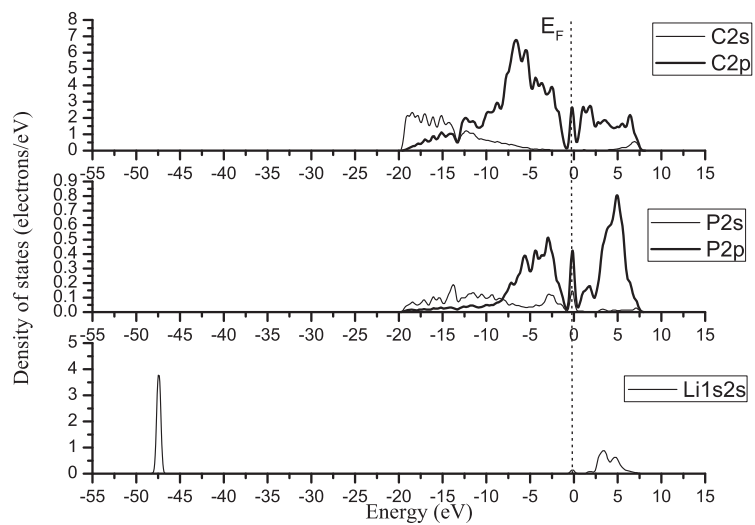}}
\caption{The partial density of states of the doped diamond where
the number of Li atom is one, P atom's number is one and the C
atom's number is fourteen. The $E_\mathrm{F}$ stands for the Fermi level. The
C$2s$ and C$2p$ represent the electric charge distributions of $2s$ and $2p$
orbits of the carbon atoms. The P$2s$ and P$2p$ represent the electric
charge distributions of $2s$ and $2p$ orbits of the phosphorus atoms.
The Li$1s2s$ represent the electric charge distributions of $1s$ and $2s$
orbits of the lithium atoms.} \label{fig5}
\end{figure}

\begin{figure}[!b]
\centerline{\includegraphics[width=0.65\textwidth]{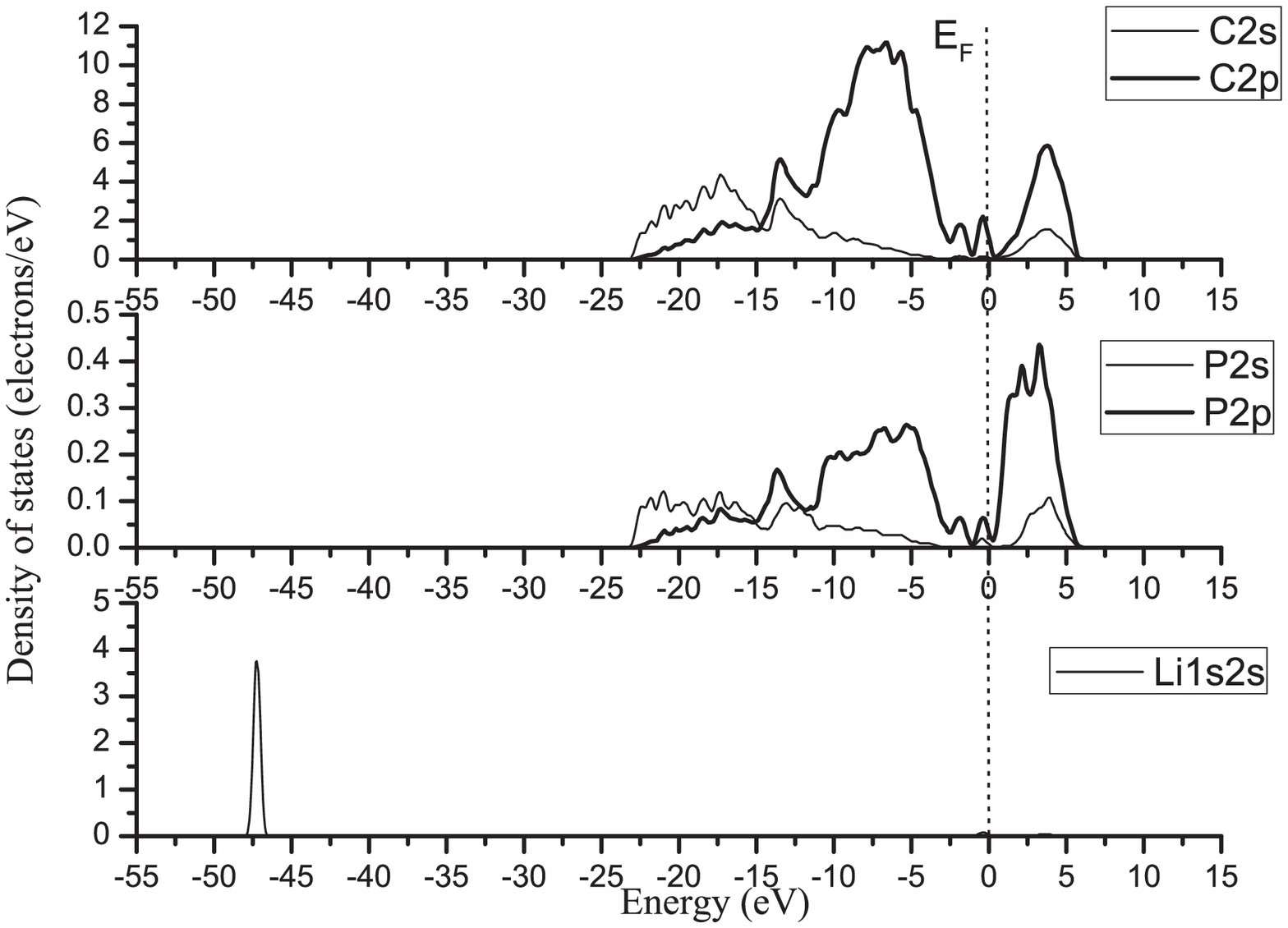}}
\caption{The partial density of states of the doped diamond where
the number of Li atom is one, P atom's number is one and the C
atom's number is thirty. The $E_\mathrm{F}$ stands for the Fermi level. The
C$2s$ and C$2p$ represent the electric charge distributions of $2s$ and $2p$
orbits of the carbon atoms. The P$2s$ and P$2p$ represent the electric
charge distributions of $2s$ and $2p$ orbits of the phosphorus atoms.
The Li$1s2s$ represent the electric charge distributions of $1s$ and $2s$
orbits of the lithium atoms.} \label{fig6}
\end{figure}

\begin{figure}[!t]
\centerline{\includegraphics[width=0.65\textwidth]{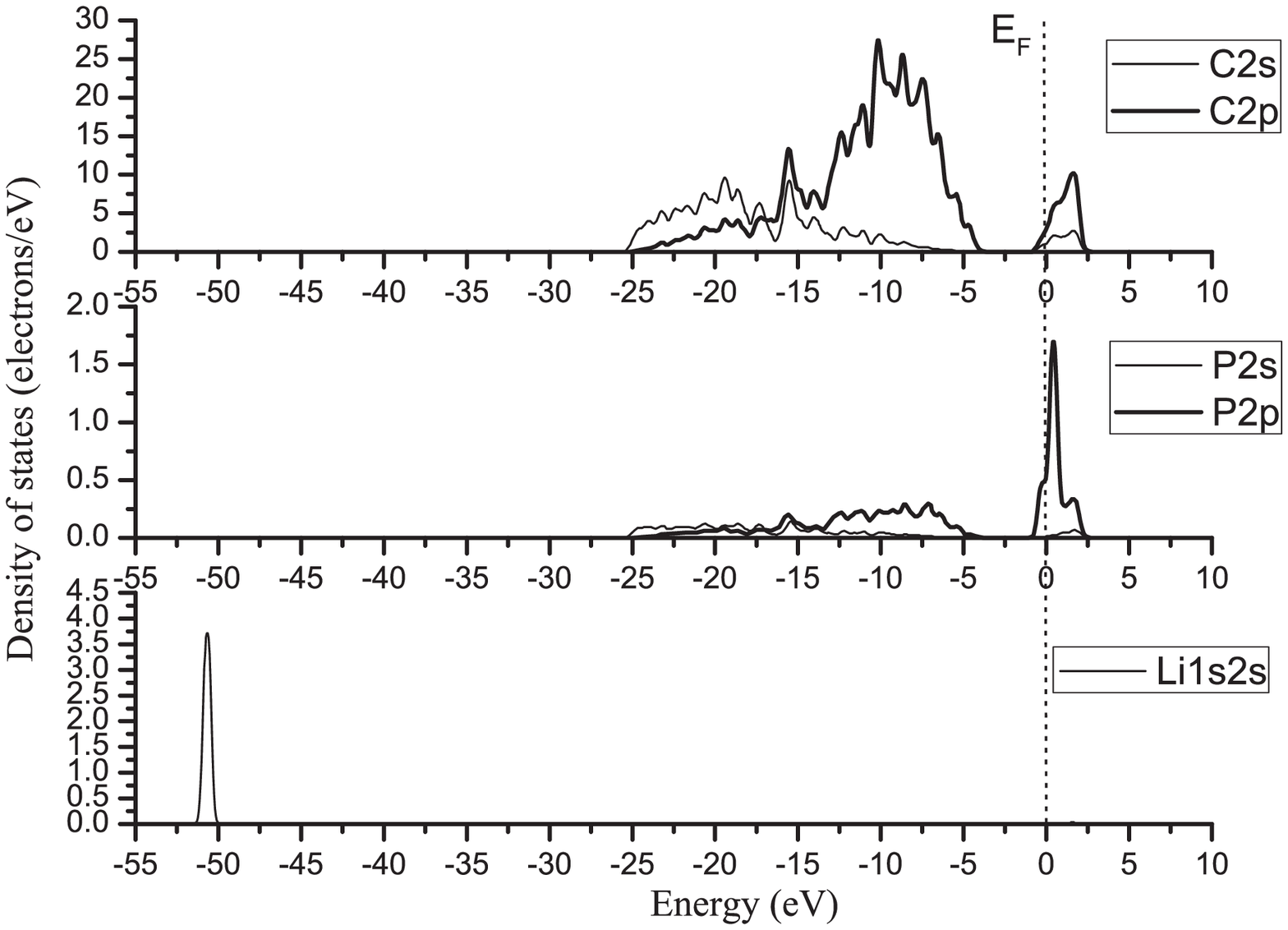}}
\caption{The partial density of states of the doped diamond where
the number of Li atom is one, P atom's number is one and the C
atom's number is sixty-two. The $E_\mathrm{F}$ stands for the Fermi level.
The C$2s$ and C$2p$ represent the electric charge distributions of $2s$
and $2p$ orbits of the carbon atoms. The P$2s$ and P$2p$ represent the
electric charge distributions of $2s$ and $2p$ orbits of the phosphorus
atoms. The Li$1s2s$ represent the electric charge distributions of $1s$
and $2s$ orbits of the lithium atoms.} \label{fig7}
\end{figure}

\begin{figure}[!b]
\centerline{\includegraphics[width=0.65\textwidth]{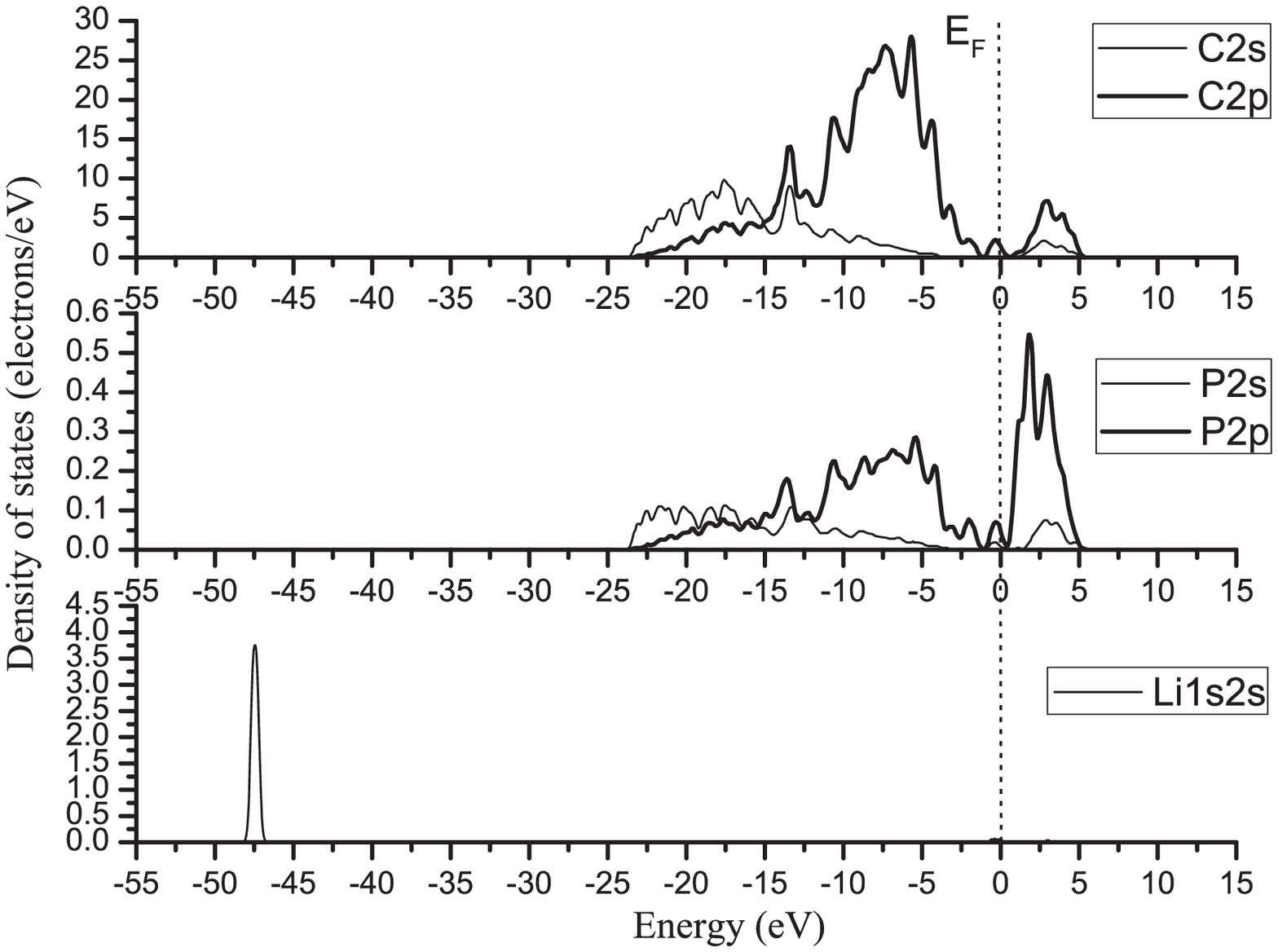}}
\caption{The partial density of states of the doped diamond where
the number of Li atom is one, P atom's number is one and the C
atom's number is seventy. The $E_\mathrm{F}$ stands for the Fermi level.
The C$2s$ and C$2p$ represent the electric charge distributions of $2s$
and $2p$ orbits of the carbon atoms. The P$2s$ and P$2p$ represent the
electric charge distributions of $2s$ and $2p$ orbits of the phosphorus
atoms. The Li$1s2s$ represent the electric charge distributions of $1s$
and $2s$ orbits of the lithium atoms.} \label{fig8}
\end{figure}

\begin{figure}[!t]
\centerline{\includegraphics[width=0.65\textwidth]{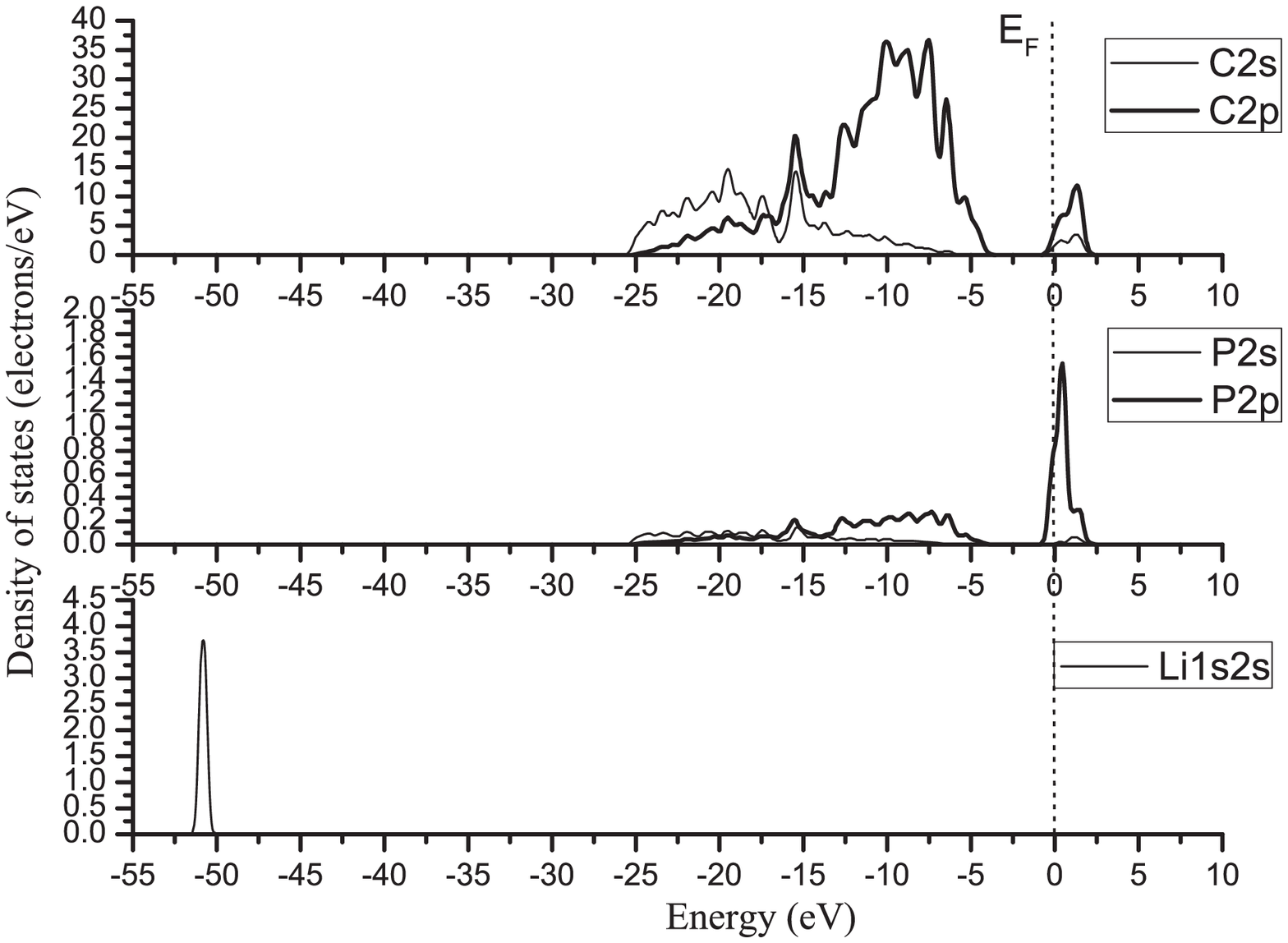}}
\caption{The partial density of states of the doped diamond where
the number of Li atom is one, P atom's number is one and the C
atom's number is ninety-four. The $E_\mathrm{F}$ stands for the Fermi
level. The C$2s$ and C$2p$ represent the electric charge distributions
of $2s$ and $2p$ orbits of the carbon atoms. The P$2s$ and P$2p$ represent
the electric charge distributions of $2s$ and $2p$ orbits of the
phosphorus atoms. The Li$1s2s$ represent the electric charge
distributions of $1s$ and $2s$ orbits of the lithium atoms.}
\label{fig9}
\end{figure}

\begin{figure}[!b]
\centerline{\includegraphics[width=0.65\textwidth]{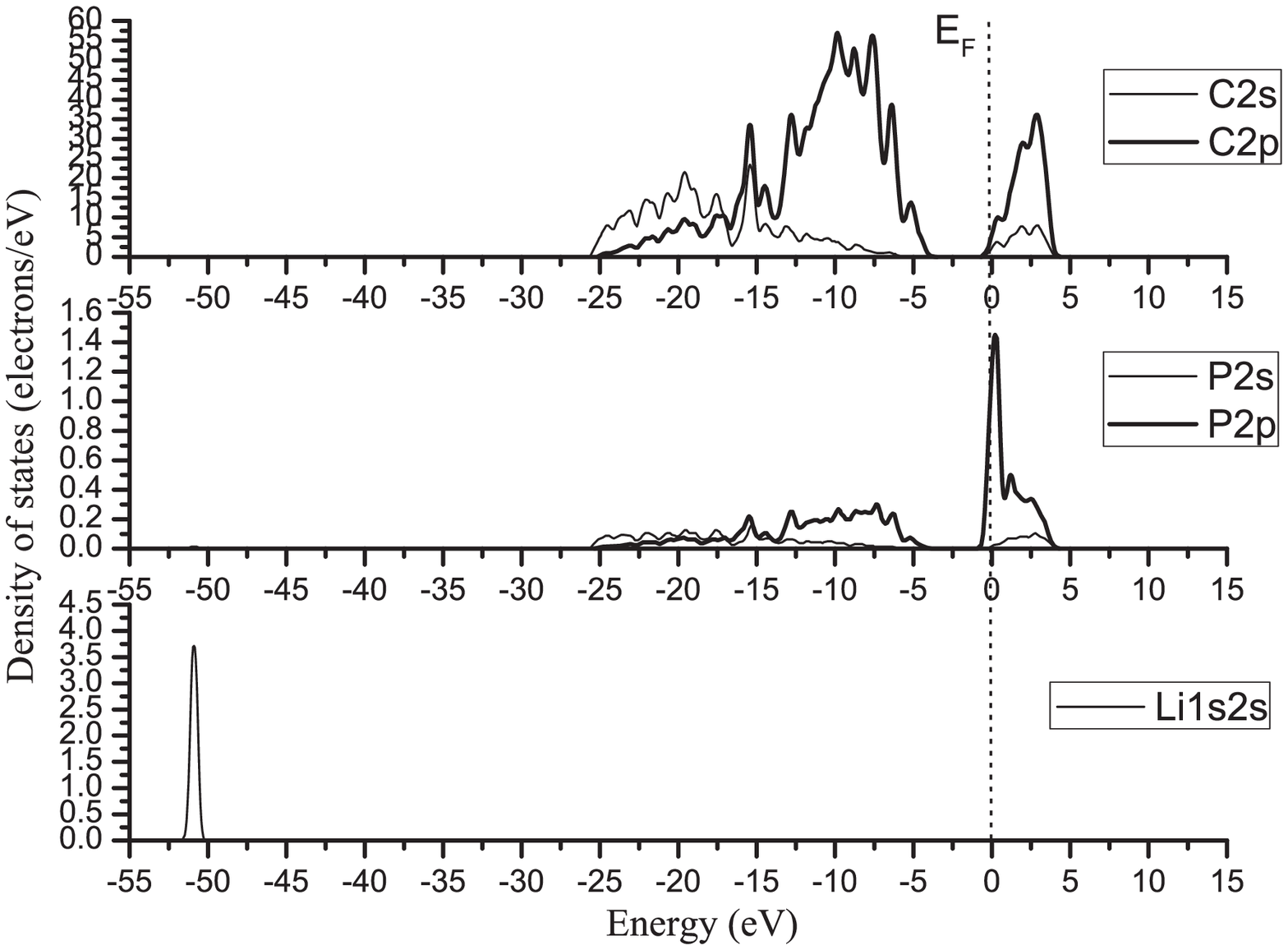}}
\caption{The partial density of states of the doped diamond where
the number of Li atom is one, P atom's number is one and the C
atom's number is one hundred and forty-two. The $E_\mathrm{F}$ stands for
the Fermi level. The C$2s$ and C$2p$ represent the electric charge
distributions of $2s$ and $2p$ orbits of the carbon atoms. The P$2s$ and
P$2p$ represent the electric charge distributions of $2s$ and $2p$ orbits
of the phosphorus atoms. The Li$1s2s$ represent the electric charge
distributions of $1s$ and $2s$ orbits of the lithium atoms.}
\label{fig10}
\end{figure}

\begin{figure}[!t]
\centerline{\includegraphics[width=0.65\textwidth]{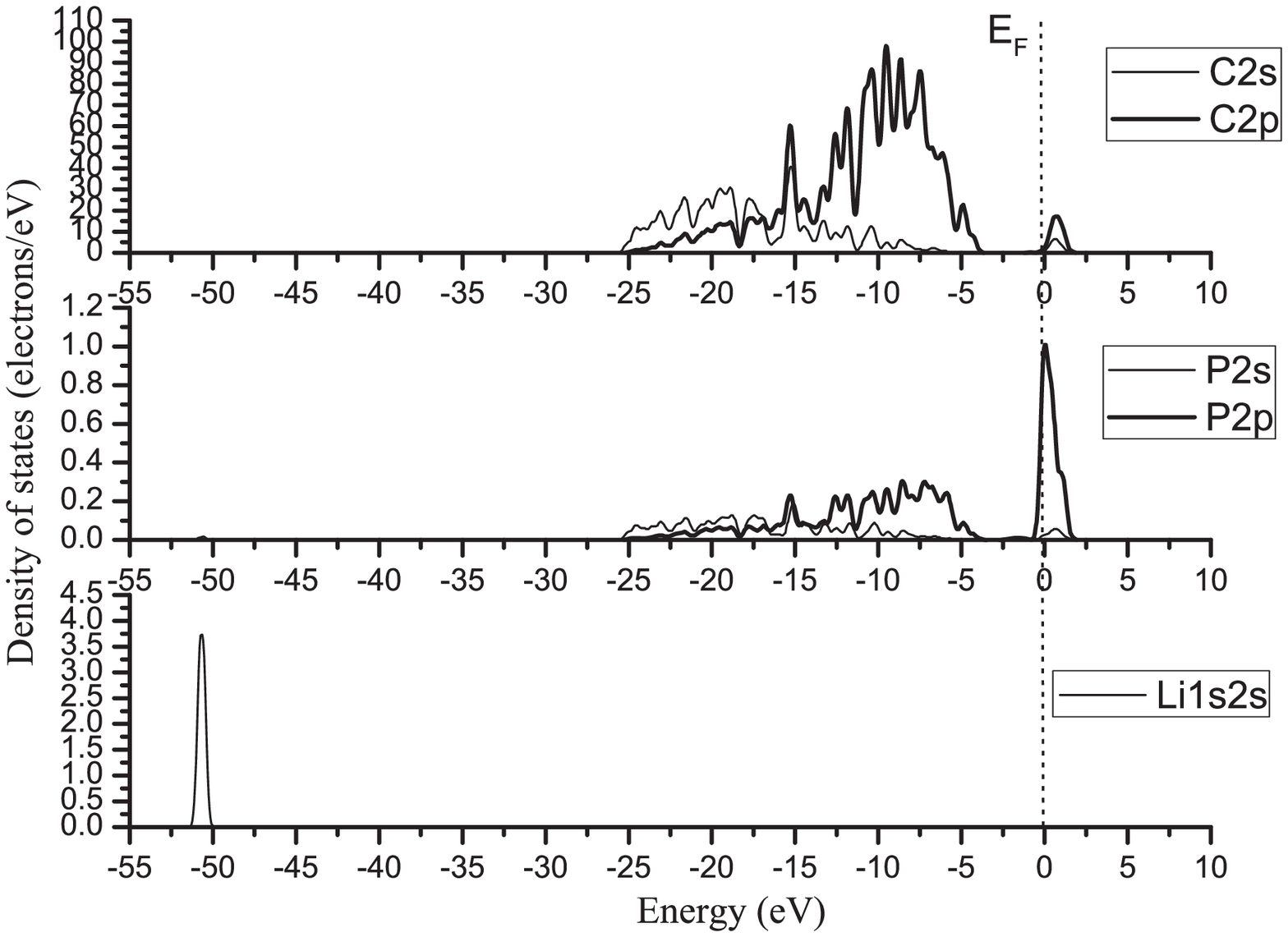}}
\caption{The partial density of states of the doped diamond where
the number of Li atom is one, P atom's number is one and the C atom's number
is two hundred and fourteen. The $E_\mathrm{F}$ stands for the Fermi level. The
C$2s$ and C$2p$ represent the electric charge distributions of $2s$ and $2p$ orbits
of the carbon atoms. The P$2s$ and P$2p$ represent the electric charge
distributions of $2s$ and $2p$ orbits of the phosphorus atoms. The Li$1s2s$
represent the electric charge distributions of $1s$ and $2s$ orbits of the
lithium atoms.} \label{fig11}
\end{figure}

In order to further analyze the impacts of the Li--P atom-pairs
co-doping on the electrical properties of the diamond, we also
calculated the PDOS of the Li--P co-doped diamond of different
doping concentrations (as shown in the figure~\ref{fig4}--\ref{fig11}). From these pictures, we can see that when the doping concentration is high (as shown in figures~\ref{fig4} and
\ref{fig5}), the $1s$ and $2s$ electrons of the lithium atom have some
contributions to the conduction band near the Fermi level; and when
the doping concentration is low (as shown in figure~\ref{fig6}--\ref{fig11}), the $s$
electrons of the lithium atom nearly have no contribution to the
conduction band near the Fermi level. Therefore, on the one hand, we
can improve the electron conductivity of the heavy phosphorus-doped
diamond by incorporating some lithium atoms; on the other hand, when
the lithium atoms are incorporated into the non-heavy
phosphorus-doped diamond, although the lithium atoms almost have no
impacts on the conductive properties of the phosphorus-doped
diamond, they can effectively reduce the vacancies and the defects of
the doped thin films, thus maintaining the integrity of the films.
In the figures, the part of the conduction band near the Fermi level
mainly comes from the contributions of the carbon atom's $2p$ orbit,
the phosphorus atom's $2p$ orbit, a little of the carbon atom's $2s$
orbit, the phosphorus atom's $2s$ orbit and the lithium atom's $1s$ and
$2s$ orbit. With different doping concentrations, the
contributions of the carbon and phosphorus atom's $2s$ orbit and the
lithium atom's $s$ orbit are different. The part of the valence
band near the Fermi level mainly comes from the contributions of the
carbon atom's $2p$ orbit and the phosphorus atom's $2p$ orbit. As the
doping concentration increases, the area of DOS of the valence band
portion will increase. Therefore, when the Li--P atom-pair is
incorporated into the diamond lattice, the covalent bond near the
Li--P atom-pair will be destroyed. Thus, under the interactions among
the outer electrons of the lithium atom, the phosphorus atom and the
carbon atoms, the energy bands of the carbon atoms and the
phosphorus atom will split near the Fermi level. The higher the
concentration, the greater the energy band splits. And the split of
the energy band near the Fermi level is propitious to improve the
conductance properties of a semiconductor.

\subsection{Orbital charge distribution of the Li--P atoms and analysis of
the bond length}

By the analysis of the Mulliken population, we can understand the
orbital electron distributions of each atom in the Li--P co-doped
diamond in detail, then to determine the bonding mechanism among
different atoms. In the first line of table~\ref{tbl1}, the $S$ (or $P$) stands for the $s$ (or $p$) orbital charge after the atom combined with
another atom, and the ``Charge'' stands for the charge an atom loses
after it combined with another atom. The orbital charge distributions
of the lithium atom and the phosphorus atom in the Li--P co-doped
diamond are showed in table~\ref{tbl1}.
\begin{table}[!t]
\caption{The orbital charge distributions of the Li and P atoms in
the Li--P co-doped diamond.}
\label{tbl1}
\begin{center}
\begin{tabular}{|c|c|c|c|c|c|}
\hline\hline
Number & atom & $S$ & $P$ & Total & Charge($e$) \\
\hline\hline
\raisebox{-1.50ex}[0cm][0cm]{8} & Li& 1.99 & 0.00 & 1.99 & 1.01 \\
\cline{2-6}
 & P & 1.49 & 2.81 & 4.30 & 0.70 \\
\hline
\raisebox{-1.50ex}[0cm][0cm]{16} & Li & 1.81 & 0.00 & 1.81 & 1.19 \\
\cline{2-6}
 & P & 1.54 & 2.67 & 4.21 & 0.79 \\
\hline
\raisebox{-1.50ex}[0cm][0cm]{32} & Li & 1.17 & 0.00 & 1.17 & 1.83 \\
\cline{2-6}
 & P & 1.21 & 2.54 & 3.75 & 1.25 \\
\hline
\raisebox{-1.50ex}[0cm][0cm]{64} & Li & 0.98 & 0.00 & 0.98 & 2.02 \\
\cline{2-6}
 & P & 1.21 & 2.63 & 3.84 & 1.16 \\
\hline
\raisebox{-1.50ex}[0cm][0cm]{72} & Li & 1.13 & 0.00 & 1.13 & 1.87 \\
\cline{2-6}
 & P & 1.21 & 2.54 & 3.75 & 1.25 \\
\hline
\raisebox{-1.50ex}[0cm][0cm]{96} & Li & 0.97 & 0.00 & 0.97 & 2.03 \\
\cline{2-6}
 & P & 1.21 & 2.67 & 3.88 & 1.12 \\
\hline
\raisebox{-1.50ex}[0cm][0cm]{144} & Li & 0.96 & 0.00 & 0.96 & 2.04 \\
\cline{2-6}
 & P & 1.21 & 2.69 & 3.90 & 1.10 \\
\hline
\raisebox{-1.50ex}[0cm][0cm]{216} & Li & 0.96 & 0.00 & 0.96 & 2.04 \\
\cline{2-6}
 & P & 1.20 & 2.79 & 3.99 & 1.01 \\
\hline\hline
\end{tabular}
\end{center}
\end{table}
In this table, both the
atomic charges of the lithium atom and the phosphorus atom are
positive, indicating that some charge of theirs has transferred to the vicinity
of the carbon atoms. When the total atomic number of a cell is
less than or equal to 32, the concentration of the impurity atoms is
high. With the concentration decrease of impurity atoms, the $s$ orbital charge number of the lithium atom decreases from 1.99 to
1.17, and the charge increases from 1.01 to 1.83 which the lithium
atom loses after it combined with another atom. This indicates that
when the doping concentration is high, then the higher is the doping
concentration of lithium atom, the less is the charge contributing to
the bonding of the lithium atom. While the total atomic number of a
cell is more than 32, the concentration of impurity atoms is low.
The $s$ orbital charge number of the lithium atom and the charge which
the lithium atom loses remain almost unchanged as the doping
concentration changes. This indicates that when the doping
concentration is low, the charge contributing to the bonding of the
lithium atom almost does not change for different doping
concentrations. As the doping concentration decreases, the $s$ orbital
charge of the phosphorus atom first increases, then decreases and
keeps an invariable value. As the doping concentration
decreases, the $p$ orbital charge of the phosphorus atom
decreases  first, and then increases. Therefore, the total change of the $s$ and
$p$ orbital charge of the phosphorus atom makes the charge which the
phosphorus loses after the bonding increases  first, and then decreases.
From the above mentioned, combined with the analysis of the DOS in
section~2.1, we can find that the interactions of the outer
electrons between the lithium atom, the phosphorus atom and the carbon
atoms are very strong. When the doping concentration is high (or the
total atomic number of a cell is less than or equal to 32), the
doping concentration has a great effect on the $s$ orbital charge
contributing to the bonding of the lithium and phosphorus atoms. When
the doping concentration is low (or the total atomic number of the
cell is more than 32), the doping concentration nearly does not
affect the $s$ orbital charge contributing to the bonding of the lithium
and phosphorus atoms, which keeps almost an invariable value.
Moreover, no matter whether in the high doping or in the low doping diamond, the
doping concentration also has a great effect on the $p$ orbital
charge contributing to bonding of phosphorus atoms. The above
results may provide some reference to doping experiments of a diamond.

\begin{table}[!t]
\caption{The bond lengths and the bond population among the nearest
neighbor Li--C atoms, Li--P atoms and the nearest neighbor P--C atoms
in the Li--P co-doped diamond.}
\label{tbl2}
\vspace{2ex}
\begin{center}
\begin{tabular}{|c|c|c|c|}
\hline\hline
Number & bond & population & length({\AA}) \\
\hline\hline
\raisebox{-3.00ex}[0cm][0cm]{8} & Li---C & $-$0.12 & 2.31932 \\
\cline{2-4}
 & Li---P & 0.12 & 2.55735 \\
\cline{2-4}
 & P---C & 0.71 & 1.75806 \\
\hline
\raisebox{-3.00ex}[0cm][0cm]{16} & Li---C & $-$0.04 & 2.34108 \\
\cline{2-4}
 & Li---P & $-$0.03 & 2.70780 \\
\cline{2-4}
 & P---C & 0.66 & 1.77178 \\
\hline
\raisebox{-3.00ex}[0cm][0cm]{32} & Li---C & $-$0.12 & 1.63001 \\
\cline{2-4}
 & Li---P & $-$0.26 & 2.09692 \\
\cline{2-4}
 & P---C & 0.66 & 1.74612 \\
\hline
\raisebox{-3.00ex}[0cm][0cm]{64} & Li---C & $-$0.32 & 1.63768 \\
\cline{2-4}
 & Li---P & $-$0.33 & 1.97483 \\
\cline{2-4}
 & P---C & 0.59 & 1.68708 \\
\hline
\raisebox{-3.00ex}[0cm][0cm]{72} & Li---C & $-$0.17 & 1.59965 \\
\cline{2-4}
 & Li---P & $-$0.29 & 2.02061 \\
\cline{2-4}
 & P---C & 0.64 & 1.72473 \\
\hline
\raisebox{-3.00ex}[0cm][0cm]{96} & Li---C & $-$0.32 & 1.62917 \\
\cline{2-4}
 & Li---P & $-$0.31 & 1.96159 \\
\cline{2-4}
 & P---C & 0.53 & 1.68730 \\
\hline
\raisebox{-3.00ex}[0cm][0cm]{144} & Li---C & $-$0.32 & 1.63096 \\
\cline{2-4}
 & Li---P & $-$0.32 & 1.94782 \\
\cline{2-4}
 & P---C & 0.53 & 1.68045 \\
\hline
\raisebox{-3.00ex}[0cm][0cm]{216} & Li---C & $-$0.32 & 1.63744 \\
\cline{2-4}
 & Li---P & $-$0.35 & 1.94659 \\
\cline{2-4}
 & P---C & 0.29 & 1.67908 \\
\hline\hline
\end{tabular}
\end{center}
\end{table}

When the lithium atom and phosphorus atom are incorporated into the
diamond, the impurity atoms can form the Li--P bond, the Li--C bonds
and the P--C bonds. The bond lengths and the bond populations of the
Li--P bond, the nearest neighbor Li--C bond and the nearest neighbor
P--C bond are showed in table~\ref{tbl2}. As the table shows, the
bond populations of the Li--C atoms are all less than zero. When
the doping concentrations of the impurity atoms are low (as the
total atomic number of the crystal cell is more than 64), the bond
population of the Li--C atoms is equal to about $-0.32$. This indicates
that the Li--C bond is anti-bonding, and as the doping
concentration decreases, the anti-bonding states reach a stable
value $-0.32$. Except that the bond population of the Li--P atoms in
the crystal cell whose total atomic number is 8 is positive, the
other bond populations of the Li--P atoms are negative, and their
values increase as the doping concentration increases. This
indicates that the Li--P bond is also anti-bonding, but the
constituents of the anti-bonding states decrease as the doping
concentration increases. When the doping concentration reaches a
certain value, the anti-bonding states of the Li--P atoms probably
transfer into the bonding states (such as the situation of the
crystal cell whose total atomic number is 8). The bond
populations of the P--C atoms are all positive, and in general their
values increase as the doping concentration increases. This
indicates that the P--C bond is the bonding, and at the same time the
constituents of the bonding states increase as the doping
concentration increases. By the bond lengths shown in
table~\ref{tbl2}, when the total atomic number of the crystal cell
is 8 or 16, the bond length of the Li--C bond is about 2.3~{\AA}, and
the average value of the Li--P atom's bond lengths is about 2.6~{\AA}. When the total atomic number of the crystal cell is more
than 32, the average value of the Li--C atom's bond lengths is about
1.6~{\AA}, the average value of the Li--P atom's bond lengths is 2.0~{\AA}. However, the bond lengths of the P-C atoms are almost uneffected by the doping concentration, and the average value of their bond
lengths is about 1.7~{\AA}. From the above analysis, the doping
concentration has a great impact on the bond lengths of the Li-C
bond and the Li--P bond, but has a little impact on the bond length
of the P--C bond.

\section{Conclusion}

By the first principle calculation theory of the DFT, in this paper
we calculate the electrical properties (such as the DOS and the
orbit charge distributions and so on) of  different doping
concentrations' Li--P co-doped diamonds. First of all, as the Li--P
atoms are incorporated into the diamond, this makes the Fermi level
of the Li--P co-doped diamond move into the vicinity of the bottom of
the conduction band and the conductance property of the Li--P
co-doped diamond thin film has been greatly improved. When the
concentration of the impurity atoms is low (the concentrations of
the lithium atom or the phosphorus atom are less than $2.35 \times
10^{21}$~cm$^{-3})$, the Li--P co-doped diamond thin film presents
the characteristic of the semiconductors. When the concentration of
the impurity atoms is high (the concentrations of the lithium atom
or the phosphorus atom are more than $2.35 \times 10^{21}$~cm$^{-3})$, the Li--P co-doped diamond thin film presents the
characteristic of the conductors. Secondly, when the doping
concentration is high, the $1s$ and $2s$ orbits of the lithium atom will
have some contributions to the conduction band near the Fermi level,
as well as may promote the splits of the phosphorus atom's and the
carbon atom's $2p$ orbits near the Fermi level. Thus, this is helpful to
improve the electron conductivity of the Li--P co-doped diamond. The
orbital charge distributions of the Li--P atoms also illustrate this
phenomenon in detail. Finally, the incorporation of the lithium atoms into
the doped diamond not only improves the electron conductivity of
semiconductors, but also may reduce the vacancies and defects of the
doped diamond thin films. At the same time, the doping concentration
of the impurity atoms also has a great impact on the bond lengths of
the Li--C atoms and the Li--P atoms. To sum up, the lithium atoms have
an effect on the electron conductivity and the integrity of the
diamond lattice in the Li--P co-doped diamond thin films.

\section*{Acknowledgements}

This work was supported by the Natural Science Foundation of Fujian Province
of China (A0220001).



\begin{thebibliography}{99}

\bibitem{Lia09} Liao~Z.H., Farrell~J.,
    J. Appl. Electrochem., 2009, \textbf{39}, 1993; \bibdoi{10.1007/s10800-009-9909-z}.
\bibitem{Luo09} Luong~J.H.T., Male~K.B., Glennon~J.D.,
    Analyst, 2009, \textbf{134}, 1965; \bibdoi{10.1039/b910206j}.
\bibitem{And10} Andrade~L.S., Rocha~R.C., Cass~Q.B., Fatibello~O.,
    Anal. Methods, 2010, \textbf{2}, 402; \bibdoi{10.1039/b9ay00092e}.
\bibitem{Yua10} Yuan~J.J., Li~H.D., Gao~S.Y., Lin~Y.H., Li~H.Y.,
    Chem. Commun., 2010, \textbf{46}, 3119; \bibdoi{10.1039/c003172k}.
\bibitem{Sop09} Sopik~B.,
    New J. Phys., 2009, \textbf{11}, 103026; \bibdoi{10.1088/1367-2630/11/10/103026}.
\bibitem{Man10} Mandal~S., Naud~C., Williams~O.A., Bustarret~E., Omnes~F., Rodiere~P., Meunier~T., Saminadayar~L., Bauerle~C.,
    Nanotechnology, 2010, \textbf{21}, 195303; \bibdoi{10.1088/0957-4484/21/19/195303}.
\bibitem{Oki10} Oki~N., Kagayama~T., Shimizu~K., Kawarada~H.,
    J. Phys. Conf. Ser., 2010, \textbf{215}, 012143; \\ \bibdoi{10.1088/1742-6596/215/1/012143}.
\bibitem{Jon96} Jones~R., Lowther~J.E., Goss~J.,
    Appl. Phys. Lett., 1996, \textbf{69}, 2489; \bibdoi{10.1063/1.117715}.
\bibitem{Koi05} Koide~Y., Koizumi~S., Kanda~H., Suzuki~M., Yoshida~H., Sakuma~N., Ono~T., Sakai~T.,
    Appl. Phys. Lett., 2005, \textbf{86}, 232105; \bibdoi{10.1063/1.1944896}.
\bibitem{Kat05} Kato~H., Futako~W., Yamasaki~S., Okushi~H.,
    Diam. Relat. Mater., 2005, \textbf{14}, 340; \bibdoi{10.1016/j.diamond.2004.11.032}.
\bibitem{Kat07} Kato~H., Yamasaki~S., Okushi~H.,
    Diam. Relat. Mater., 2005, \textbf{14}, 2007; \bibdoi{10.1016/j.diamond.2005.08.021}.
\bibitem{Yam06} Yamada~T., Okano~K., Yamaguchi~H., Kato~H., Shikata~S., Nebel~C.E.,
    Appl. Phys. Lett., 2006, \textbf{88}, 212114; \bibdoi{10.1063/1.2206552}.
\bibitem{Per08} Pernot~J., Koizumi~S.,
    Appl. Phys. Lett., 2008,\textbf{ 93}, 052105; \bibdoi{10.1063/1.2969066}.
\bibitem{Cao95} Cao~G.Z., Driessen~F.A.J.M., Bauhuis~G.J., Giling~L.J., Alkemade~P.F.A.,
    J. Appl. Phys., 1995, \textbf{78}, 3125; \\ \bibdoi{10.1063/1.359998}.
\bibitem{LiR04} Li~R.B., Hu~X.J., Shen~H.S., He~X.C.,
    J. Mater. Sci., 2004, \textbf{39}, 1135; \bibdoi{10.1023/B:JMSC.0000012963.04082.a1}.
\bibitem{Lee05} Lee~W.S., Yu~J., Lee~T.Y.,
    J. Mater. Sci., 2005, \textbf{40}, 5549; \bibdoi{10.1007/s10853-005-4548-1}.
\bibitem{Lom07} Lombardi~E.B., Mainwood~A., Osuch~K.,
    Phys. Rev. B, 2007, \textbf{76}, 155203; \bibdoi{10.1103/PhysRevB.76.155203}.
\bibitem{Ior09} Iori~F., Ossicini~S.,
    Physica E, 2009, \textbf{41}, 939; \bibdoi{10.1016/j.physe.2008.08.010}.
\bibitem{Seg02} Segall~M.D., Lindan~P.J.D, Probert~M.J.J.,
    J. Phys.: Condens. Matter, 2002, \textbf{14}, 2717; \\\bibdoi{10.1088/0953-8984/14/11/301}.
\bibitem{Van90} Vanderbilt~D.,
    Phys. Rev. B, 1990, \textbf{41}, 7892; \bibdoi{10.1103/PhysRevB.41.7892}.
\bibitem{Per96} Perdew~J.P., Burke~K., Ernzerhof~M.,
    Phys. Rev. Lett., 1996, \textbf{77}, 3865; \bibdoi{10.1103/PhysRevLett.77.3865}.
\bibitem{Chi89} Ching~W.Y., Xu~Y.N., Wong~K.W.,
    Phys. Rev. B, 1989, \textbf{40}, 7684; \bibdoi{10.1103/PhysRevB.40.7684}.
\end{thebibliography}

%

\newpage

\ukrainianpart

\title{Першопринципні розрахунки літієво-фосфорного співлегованого алмазу}

\author{К.І. Шао\refaddr{label1,label2}, Г.У. Ванг\refaddr{label1}, Дж. Жанг\refaddr{label1}, К.Г. Жу\refaddr{label3}}

\addresses{
\addr{label1} Лабораторія квантово-інформаційних технологій, Школа фізики та телекомунікаційних технологій, \\
Педагогічний університет Південного Китаю, Гуанчжоу 510006, Китай
\addr{label2} Фізичний факультет, Педагогічний університет Жангжоу, Жангжоу 363000, Китай
\addr{label3} Фізичний факультет,  Бейханський університет, Пекін 100191, Китай
}

\makeukrtitle

\begin{abstract}
\tolerance=3000%
Ми обчислюємо густину станів (DOS) і заселення Муллікена алмазу та співлегованих алмазів з різними концентраціями літію (Li) і фосфору (P) за допомогою методу функціоналу густини та аналізуємо випадки зв'язування тонких плівок Li--P співлегованого алмазу, а також впливи Li--P співлегування на провідність алмазу. Результати показують, що атоми Li--P можуть активізувати розщеплення енергетичної зони алмазу поблизу рівня Фермі, а отже покращити провідність електронів тонких плівок Li--P співлегованого алмазу, або ж навіть перетворити Li--P співлегований алмаз з напівпровідника у провідник. Проаналізовано вплив Li--P концентрації співлегування на орбітальний розподіл заряду, довжину зв'язку та заселеність зв'язку. Атом Li може активізувати розщеплення енергетичної зони поблизу рівня Фермі, а також може благотворно регулювати спотворення та розширення кристалічної гратки алмазу.

\keywords  Li--P співлегований алмаз, густина станів, рівень домішок, орбітальний заряд
\end{abstract}

\lastpage
\end{document}